\documentclass[12pt]{article}
\usepackage{amssymb}

\usepackage{natbib,graphicx,setspace}
\usepackage{amsmath,amsthm,bm,color}

\usepackage{xcolor}         
\usepackage{fullpage,amsmath}                        
\usepackage{caption}
\usepackage{hyperref}
\usepackage[rightbars, color]{changebar}
\definecolor{DarkGreen}{rgb}{0.5,0.8,0.6}   
\definecolor{RGBblack}{rgb}{0.0,0.0,0.0}    
                                          
\newtheorem{theorem}{Theorem}[section]

\newenvironment{definition}[1][Definition]{\begin{trivlist}
\item[\hskip \labelsep {\bfseries #1}]}{\end{trivlist}}

\def\hei{\color{black}}

\newcommand{\ech}{\hei \rm}

\newcommand{\is}{\itemsep=0pt}
\newcommand{\bd}[1]{\begin{description}[#1]\is}
  \newcommand{\ed}{\end{description}}
\newcommand{\bi}{\begin{itemize}\is}
  \newcommand{\ei}{\end{itemize}}
\newcommand{\be}{\begin{enumerate}\is}
  \newcommand{\ee}{\end{enumerate}}
  \newcommand{\beq}{\begin{eqnarray}\is}
  \newcommand{\eeq}{\end{eqnarray}}
\makeatletter
\newcommand*{\rom}[1]{\expandafter\@slowromancap\romannumeral #1@}

\newcommand{\bx}{\bm{x}}

\newcommand{\by}{\bm{y}}
\newcommand{\bn}{\bm{n}}
\newcommand{\bN}{\bm{N}}

\newcommand{\bZ}{\bm{Z}}
\newcommand{\bpi}{\bm{\pi}}
\newcommand{\tbZ}{\bm{\tilde{Z}}}

\newcommand{\bz}{\bm{z}}
\newcommand{\bw}{\bm{w}}
\newcommand{\bp}{\bm{p}}

\newcommand{\sig}{\sigma}

\newcommand{\Chat}{\widehat{C}}
\newcommand{\Zhat}{\widehat{\bZ}}
\newcommand{\zhat}{\widehat{z}}

\newcommand{\what}{\widehat{w}}
\newcommand{\pbar}{\bar{p}}

\newcommand{\bZt}{\widetilde{\bZ}}
\newcommand{\zt}{\widetilde{z}}

\newcommand{\teta}{\widetilde{\eta}}
\newcommand{\tmu}{\widetilde{\mu}}
\newcommand{\tphi}{\widetilde{\phi}}

\newcommand{\tpsi}{\widetilde{\psi}}
\newcommand{\tsig}{\widetilde{\sig}}
\newcommand{\tp}{\widetilde{p}}

\newcommand{\Dir}{\mbox{Dir}}

\newcommand{\Binom}{\mbox{Bin}}
\newcommand{\Poi}{\mbox{Poi}}
\newcommand{\Bin}{\mbox{Bin}}
\newcommand{\TRUE}{o}

\newcommand{\Ks}{C}

\begin{document}
\doublespacing

\title{MAD Bayes for Tumor Heterogeneity --
Feature Allocation with Exponential Family Sampling} 
\author{
            Yanxun Xu\\
            {\small Division of Statistics and Scientific Computing, The University of Texas at Austin, Austin, TX}
            \and Peter M\"uller \thanks{Address for Correspondence:  Department of Mathematics
UT Austin 1, University Station, C1200, Austin, TX 78712 USA. E-mail: pmueller@math.utexas.edu. jiyuan@uchicago.edu. }\\
            {\small Department of Mathematics, The University of Texas at Austin, Austin, TX} 
        \and Yuan Yuan\\
        {\small SCBMB, Baylor College of Medicine, Houston, TX}
        \and Kamalakar Gulukota\\
        {\small Center for Biomedical Informatics, NorthShore University
          HealthSystem, Evanston, IL}\\
          \and Yuan Ji $^{*}$\\
          {\small   Center for Biomedical Informatics, NorthShore University
          HealthSystem, Evanston, IL}\\
          {\small Department of Health Studies, The University of Chicago, Chicago, IL}          
}           
\date{}
\maketitle

\clearpage
\newpage

\begin{abstract}
We propose small-variance asymptotic approximations for inference
on tumor heterogeneity (TH) using next-generation sequencing data.
Understanding TH is an important and open research problem in
biology. The lack of appropriate statistical inference is a
critical gap in existing methods that the proposed approach aims to
fill. 
We build on a hierarchical model with an exponential family likelihood
and a feature allocation prior. 
The proposed   implementation of posterior inference  
generalizes similar small-variance
approximations proposed by Kulis and Jordan (2012) and
Broderick et.al (2012b) for inference with Dirichlet process mixture
and Indian buffet process prior models under normal sampling.
We show that the new algorithm can successfully recover
latent structures of different haplotypes and 
subclones and is magnitudes faster
than available Markov chain Monte Carlo samplers.
  The latter are   practically infeasible for high-dimensional genomics data. 
The proposed approach is scalable, easy to implement and
benefits from the flexibility of Bayesian nonparametric models.  
More importantly, it provides a useful tool for 
  applied scientists  
to estimate cell subtypes in tumor samples. R code is available on
\href{http://www.ma.utexas.edu/users/yxu/}{http://www.ma.utexas.edu/users/yxu/}.

\end{abstract}

\section{Introduction}
\label{sec:intro}
\subsection{MAD-Bayes}
We propose a generalization of the MAD (maximum a posteriori
based asymptotic derivations) Bayes approach of
\cite{broderick2012mad} to latent 
feature models beyond the conjugate normal-normal setup. 
The model is developed for inference on tumor heterogeneity (TH), when
the sampling model is a binomial distribution for observed  short
reads counts for 
single nucleotide variants (SNVs)  in next-generation sequencing
(NGS) experiments. 

The proposed model includes a Bayesian non-parametric (BNP)
prior. 
BNP models are characterized by parameters that live on an
infinite-dimensional parameter space, such as unknown mean functions
or unknown probability measures.
Related methods are widely used in a variety of machine learning and
biomedical research problems, including clustering, regression and
feature allocation.
While BNP methods are flexible from a modeling perspective,
a major limitation is the computational challenge that arises in
posterior inference with large-scale problems and big data.  Posterior
inference in highly structured models is often implemented by Markov chain
Monte Carlo (MCMC) simulation \citep[for example]{liu2008monte}
or variational inference such as
expectation-maximization (EM) algorithm \citep{dempster1977maximum}.
However, neither approach scales effectively to high-dimensional data.
As a result, simple {\it ad-hoc} methods, such as K-means
\citep{hartigan1979algorithm}, are still preferred in many large-scale
applications.
K-means clustering is often preferred over full posterior
inference in model-based clustering, such as Dirichlet process (DP)
mixture models. DP mixture models are some of the most widely used BNP models. 
See, for example, \cite{Ghoshal:10}, for a review.
 
Despite the simplicity and scalability, K-means has some
known shortcomings. 
First, the K-means algorithm is a rule-based method. 
The output is an {\it ad-hoc} 
point estimate of the unknown partition. There is no notion of
characterizing uncertainty, and it is difficult to coherently embed it
in a larger model. 
Second, the K-means algorithm requires a fixed number of
clusters, which is not available in many applications.  An ideal
algorithm should combine the scalability of K-means with the
flexibility of Bayesian nonparametric models. Such links
between non-probabilistic (i.e., rule-based methods like
K-means) 
and probabilistic approaches (e.g., posterior MCMC or the EM
algorithm)  can sometimes be found by using small-variance
asymptotics.
For example, the EM algorithm for a mixture of Gaussian model
becomes the K-means algorithm as the variances of the
Gaussians tend to zero \citep{trevor2001elements}. In general,
small-variance asymptotics can offer useful alternative approximate
implementations of inference for large-scale Bayesian nonparametric
models, exploiting the fact that corresponding non-probabilistic
models show advantageous scaling properties.

Using small-variance asymptotics, \cite{kulis2011revisiting}
showed how a K-means-like algorithm could approximate
posterior inference 
for Dirichlet process (DP) mixtures.  
\cite{broderick2012mad} generalized the approach by developing 
small-variance asymptotics to MAP (maximum a posteriori)
estimation in feature allocation models with Indian buffet process
(IBP) priors \citep{griffiths2005infinite, teh2007stick}. Similar
to the K-means algorithm, they proposed the BP (beta process)-means algorithm for
feature learning. 
 Both approaches are 
restricted to normal sampling and conjugate normal priors, which
 facilitates the asymptotic argument and
greatly simplifies the computation. However, it is not immediately generalizable
to other distributions, preventing their methodology from being
applied to non-Gaussian data. 
The application that motivates the current paper is a typical
example.  We require posterior inference for a feature allocation
model with a binomial sampling model.

\subsection{Tumor Heterogeneity}
The proposed methods are motivated
by an application to inference for tumor heterogeneity (TH). This is a
highly important and open research
problem that is currently studied by many cancer
researchers \citep{gerlinger2012intratumor, landau2013evolution, larson2013purbayes, andor2014expands, roth2014pyclone}. 
In the literature over the past five years a consensus has emerged 
that 
tumor cells are heterogenous, both within the same biological tissue sample
and between different samples.  A tumor sample typically comprises an
 admixture of subtypes of different cells, each possessing a
unique genome. We will use the term ``subclones'' to 
refer to cell
subtypes in a biological sample.   Inference on   genotypic differences
(differences in DNA base pairs) between subclones and
proportions of each subclone in a sample   can provide critical   new
information for cancer diagnosis and
prognosis. However, inference and statistical modeling are challenging
and few solutions exist.

\begin{figure}[!]
\centering
\begin{tabular}{c}
\includegraphics[width=.75\textwidth]{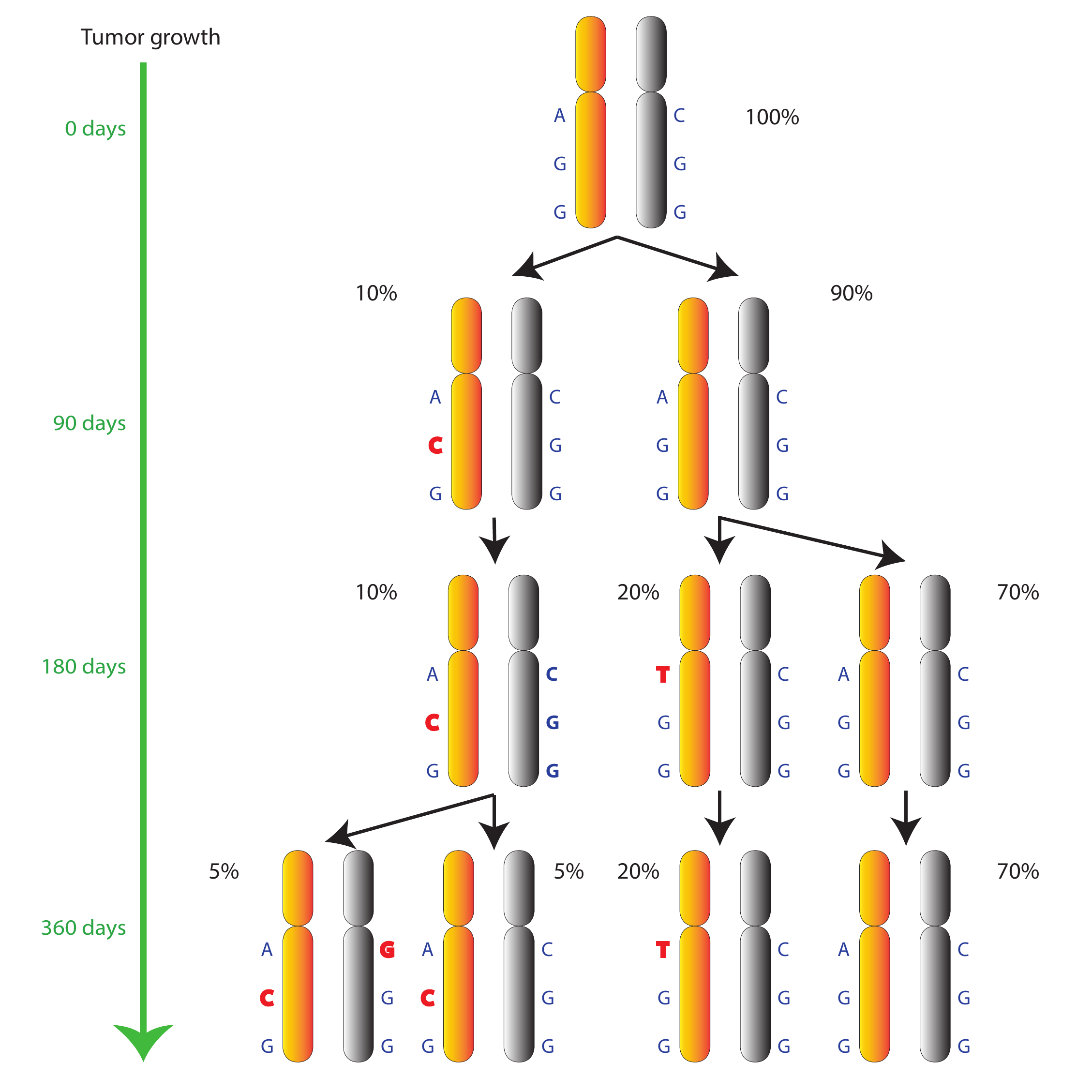}\\
(a)\\
\includegraphics[width=.75\textwidth]{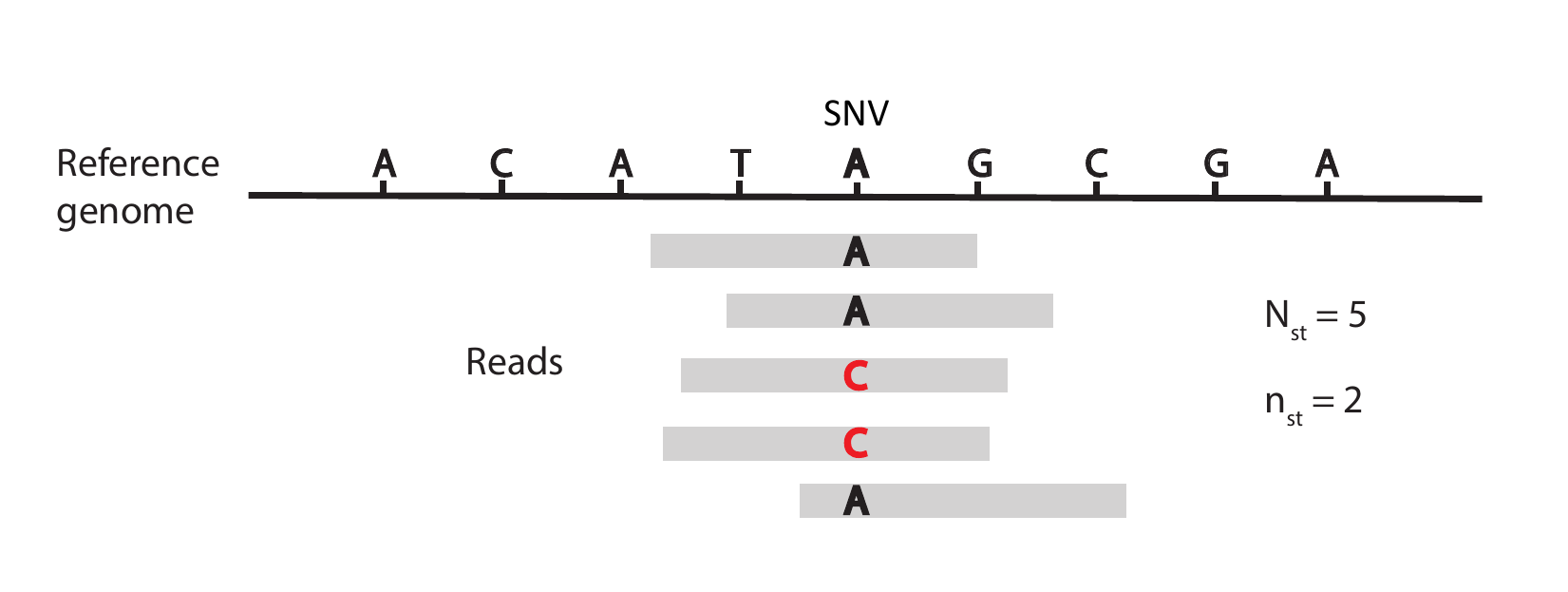}\\
(b)\\
\end{tabular}
\caption{(a) Somatic mutations (in red) that are acquired during tumor
  growth lead to the emergence of new subclones and haplotypes. (b) An
  illustration of an SNV and read mapping. The figure shows five short
  reads that are mapped to the indicated SNV location. The reference
  genotype is A. Among the five reads, two reads have a variant
  genotype C. 
  So $N_{st}=5$, $n_{st}=2$ and the observed VAF is $2/5$. }
\label{fig:mapping}
\end{figure}

Genotypic differences between subclones do not occur frequently. They
are often restricted to single nucleotide variations
(SNVs). When a sample is heterogeneous, it contains multiple
subclones  with each subclone  possessing a unique
genome. Often the differences 
between subclonal genomes are marked by somatically acquired SNVs. For
multiple samples from the same tumor, intra-tumor heterogeneity refers to the
presence of multiple subclones  that appear in different proportions across
different samples. 
For samples from different patients, however,
subclonal genomes are rarely shared due to polymorphism between
patients. However, for a selected set of potentially disease related SNVs
(e.g., from biomarker genes), one
may still find locally shared haplotypes  (a set of SNVs on the same chromosome)
across patients consisting of the selected SNVs. 
Importantly,   in the upcoming discussion  
we regard two subclonal genomes the same if they possess identical
genotypes on the   selected   SNVs, regardless of the rest of
genome. In other 
words, we do not insist that the whole genomes of any two cells must
be identical in order to call them subclonal. 
Figure
\ref{fig:mapping}(a) illustrates how such different
subclones 
can develop over the life history of a tumor, ending, in
this illustration, with three subclones and five unique haplotypes
(ACG, GGG, CGG, TGG, AGG). 

We start with models for haplotypes. Having more than two
haplotypes in a sample implies cellular heterogeneity. 
We use NGS data that record for each sample $t$ the   number
$N_{st}$ of short reads that are mapped 
to the genomic location of each of the selected SNVs $s$, $s=1,\ldots,S$. 
Some of these overlapping reads possess a variant sequence,
while others include the reference sequence.  
Let $n_{st}$ denote the number of variant reads among the $N_{st}$
overlapping reads for each sample and SNV.  
In Figure
\ref{fig:mapping}(b),
$N_{st}=5$ reads are mapped to the indicated SNV. While the
reference sequence is ``A'', $n_{st}=2$
short reads bear a variant sequence ``C''. 
We define variant allele fraction (VAF) as the proportion
$n_{st}/N_{st}$ of short reads
bearing a variant genotype among all the short reads that are mapped to an
SNV. In Figure \ref{fig:mapping}(b), the VAF is 2/5 for that SNV. In
Section 2, we will model the observed variant read counts by 
a mixture of latent haplotypes, which in turn  
are defined by a pattern of present and absent SNVs.
Assuming that each sample is composed of some proportions of these
haplotypes, we can then fit the observed VAFs across SNVs in each
sample.
Formally, modeling involves binomial sampling models 
for the observed counts with mixture priors for the binomial
success probabilities. The mixture is over an (unknown) number $C$ of
(latent) haplotypes, which in turn are
represented as a binary matrix $\bZ$, with
columns, $c=1,\ldots,C,$ corresponding to haplotypes and rows,
$s=1,\ldots,S$, corresponding to SNVs. The entries $z_{sc} \in \{0,1\}$ are
indicators for variant allele $s$ appearing in haplotype
$c$. That is, each column of indicators defines a haplotype
 by specifying the 
genotypes (variant or not) of the corresponding  SNVs. 

\subsection{Main Contributions}  
A key element of the proposed model is the prior on the binary matrix
$\bZ$.  
We recognize the problem as a special case of a feature
allocation problem and use an Indian buffet process (IBP) prior for $\bZ$.
In the language of feature
allocation models (and the traditional metaphor that is used to
describe the IBP prior),
the haplotypes are the features (or dishes)  and the SNVs are
experimental units (or customers) that select features. Each tumor
sample consists of 
an unknown proportion of these haplotypes.  
\cite{juhee2013feature} used a finite feature allocation model to
describe the latent structure of possible haplotypes.
The model is restricted to a
fixed number of haplotypes. In practice, the number of
haplotypes is unknown. A possible way to 
generalize to an
unknown number of
latent features is to define a transdimensional MCMC scheme, such as a
reversible jump (RJ) algorithm \citep{green1995reversible}. However,
it is difficult to implement a practicable RJ algorithm.

An attractive alternative is to 
generalize the BP-means algorithm of 
\cite{broderick2012mad}
beyond the Gaussian case. Inspired by the connection between Bregman
divergences \citep{bregman1967relaxation} and exponential families, we
propose a MAP-based small-variance asymptotic approximation for any
 exponential family likelihood with an IBP feature
allocation prior. We call the proposed approach  the FL-means
algorithm, where FL stands for ``feature learning''. 
The FL-means algorithm is scalable, easy to implement and benefits from
the flexibility of Bayesian nonparametric models. 
  Computation time is reduced beyond a factor 1,000.  

This paper proceeds as follows. In Section \ref{sec:modelTH}, we
introduce the Bayesian feature allocation model to describe tumor
heterogeneity. Section \ref{sec:MAD} elaborates the FL-means
algorithm and proves convergence. We examine the
performance of FL-means through simulation studies in Section
\ref{sec:simu}.  In Section \ref{sec:analysis}, we apply the FL-means
algorithm to inference for TH. Finally, we conclude
with a brief discussion in Section \ref{sec:con}.

\section{A Model for Tumor Heterogeneity}
\label{sec:modelTH}
\subsection{Likelihood}
We consider data sets from NGS experiments. 
The data sets record the observed read counts 
for $S$ SNVs  in $T$ tumor samples.
Let $\bn$ and $\bN$ denote $S \times T$ matrices with
$N_{st}$ denoting the total number of reads overlapping with SNV $s$
in tissue sample $t$, and 
$n_{st}$ denoting the number of variant reads among those $N_{st}$
reads. 
The ratio $n_{st}/N_{st}$ is the observed VAF. Figure
\ref{fig:mapping}(b) provides an illustration.  
We assume a binomial sampling model 
$$
   p(n_{st} \mid N_{st},p_{st}) = \Binom(N_{st}, p_{st}),
$$
where $p_{st}$ is the expected VAF, $p_{st} =
E(n_{st}/N_{st} \mid N_{st},p_{st})$.
Conditional on $N_{st}$ and $p_{st}$, the observed counts $n_{st}$ are
independent across $s$ and $t$. The likelihood becomes
\begin{equation}
  p(\bn\mid \bN, \bp) = \prod_{s=1}^S\prod_{t=1}^T 
     {N_{st} \choose
       n_{st}}p_{st}^{n_{st}}(1-p_{st})^{(N_{st}-n_{st})},
\label{eq:like}
\end{equation}
where $\bn=[n_{st}]$, 
$\bN=[N_{st}]$ and
$\bp=[p_{st}]$ are $(S \times T)$ matrices. 
The binomial success probabilities $p_{st}$ are modeled in terms of
latent haplotypes which we introduce next.

\subsection{Prior Model for Haplotypes}
\begin{figure}[h!]
  \centering{
    \includegraphics[width=.3\textwidth]{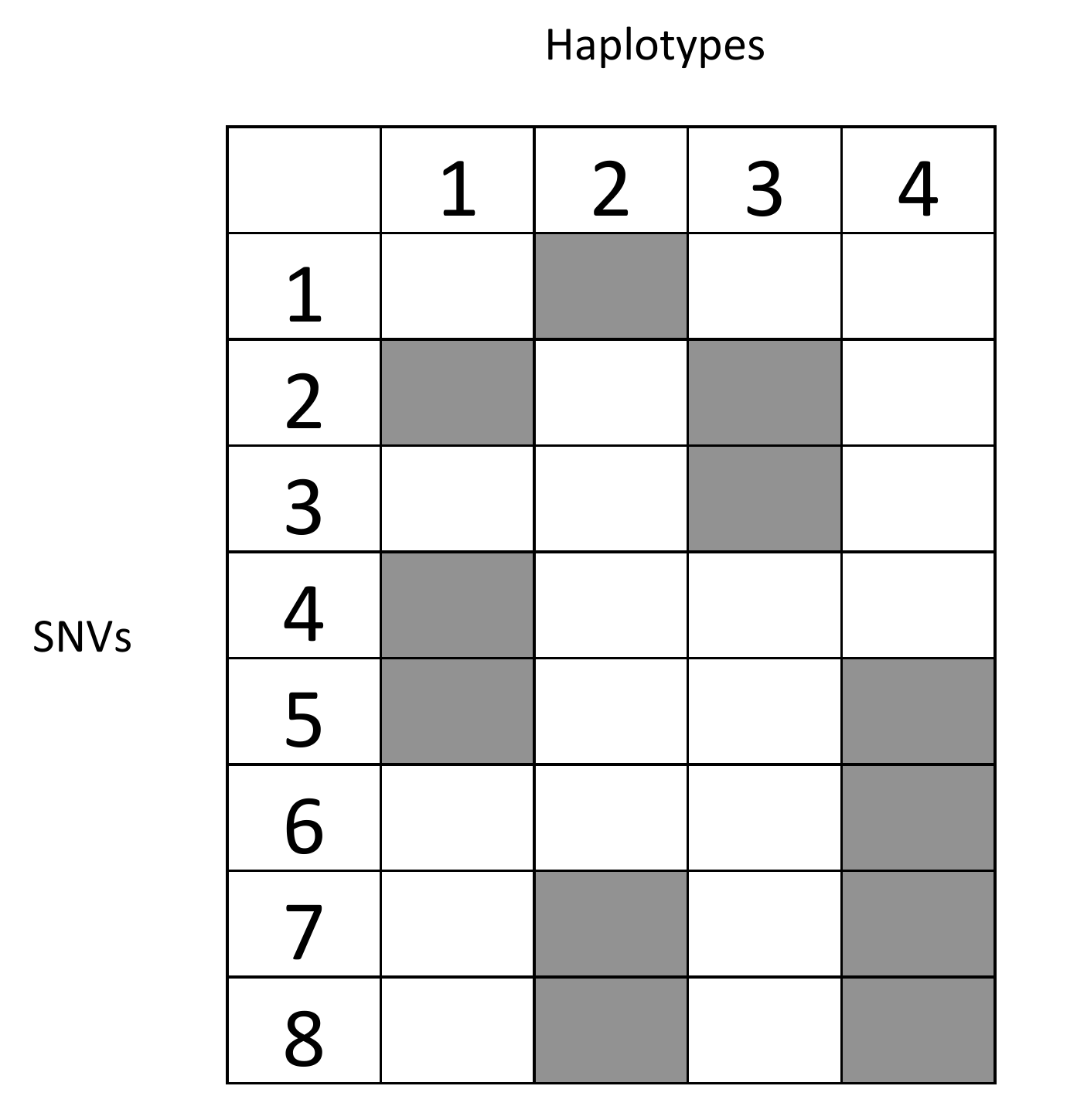}}
  \caption{A binary matrix $\bZ$ with haplotypes (latent
    features) $c=1,\ldots,C(=4)$, in the columns
    and $S=8$ SNVs in the rows.
    A shaded cell indicates that an SNV occurs in a haplotype
    ($z_{sc}=1$). }
  \label{fig:illustration}
\end{figure}

Recall that $\{z_{sc} =1\}$ indicates that the genotype at SNV $s$
is a variant (different from the reference genotype) for haplotype $c$. 
Figure \ref{fig:illustration} illustrates the
binary latent matrix $\bZ=[z_{sc}]$ with $C=4$ haplotypes (columns)
and $S=8$ SNVs (rows). A shaded cell indicates $z_{sc}=1$. For instance,
SNV 2 occurs in two haplotypes $c=1$ and 3, SNV 3 occurs only in
haplotype $c=3$. 
Assuming that each sample is composed of proportions 
$w_{tc}$ of the haplotypes, $c=1,\ldots,C$,
we represent $p_{st}$
as
\begin{eqnarray}
    p_{st} = w_{t0}p_0 + \sum_{c=1}^{C} w_{tc} z_{sc},
    \label{eq:model_p}
\end{eqnarray}
where $w_{tc} \in [0,1)$ and $\sum_{c=0}^C w_{tc} = 1$. 
 
In \eqref{eq:model_p}, we introduce an additional background
haplotype $c=0$ that captures experimental and data processing noise,
such as the fraction of variant reads that are mapped with low quality or error. 
In the decomposition, $p_0$ is the relative frequency of an SNV in the background.
 Equation \eqref{eq:model_p} is a key model assumption. It allows us
to deduce the unknown haplotypes from a decomposition of the 
expected VAF $p_{st}$ as a weighted sum of latent genotype calls $z_{sc}$
with weights being the proportions of haplotypes. 
  Importantly, we assume these weights to be the same across all
SNVs, $s=1,\ldots,S$.  
In other words, the expected VAF is contributed by those haplotypes
with variant genotypes, weighted by the haplotype prevalences.
Haplotypes without variant genotype on SNV $s$ do not contribute to
the VAF for $s$ since all the short reads generated from those
haplotypes are normal reads. 
For example, in Figure
\ref{fig:illustration} the variant reads mapped to SNV 2 should come
from haplotypes 1 and 3, but not from 2 or 4.  

The next step in the model construction is the prior model $p(\bZ$).
We use the IBP prior. See, for example
\cite{griffiths2005infinite},
\cite{thibeauxjordan:07}
or \cite{teh2007stick}
for a discussion of the IBP and for a generative model. 
We briefly review the construction in the context of the current
application. We build the binary matrix $\bZ$ one line (SNV) at a time,
adding columns (haplotypes) when the first SNV appears with $z_{sc}=1$.
Let $C_s$ denote the number of columns that are constructed by the
first $s$ SNVs. 
For each SNV, $s=1,\ldots,S$, we introduce $\Ks_s^+ \sim \Poi(\gamma/s)$ new
haplotypes, and assign $z_{sc} = 1$ and $z_{s',c}=0, \forall s'<s$,
for the new haplotypes $c=\Ks_{s-1}+1,\ldots,\Ks_{s-1}+\Ks_s^+=\Ks_s.$
For the earlier haplotypes, $c=1,\ldots,\Ks_{s-1}$, mutation $s$ is
included with probability $p(z_{sc}=1)=\frac{m_{sc}}{s}$ (note the
$s$, rather than $s-1$ in the denominator), where $m_{sc}=\sum_{s'<s}
z_{s'c}$ denote the column sums up to row $s-1$, that is, the number
of SNVs, $s'<s$, that defines haplotype $c$.
 Implicit in the construction is a convention of indexing columns
by order of appearance, that is, as columns are added for SNVs,
$s=1,\ldots,S$.
While it is customary to restrict $\bZ$ to this so called left-ordered 
form, we use a variation of the IBP without this constraint, discussed
in \cite{griffiths2005infinite} or \cite{broderick2012beta}. 
After removing the order constraint, that is, with uniform permutation
of the column indices, we get the IBP prior for a binary matrix, {\em without
  left order constraint},
\begin{equation}
  p(\bZ \mid \gamma) =
  \frac{\gamma^C\, e^{-\gamma H_S}}
       {C!}
\;
  \prod_{c=1}^C
  \frac{(S-m_c)!(m_c-1)!}
       {S!},
\label{eq:IBP1}
\end{equation}
with a random number of columns (haplotypes) $C$ and a fixed number of
rows (SNVs) $S$. Here $m_c$ denote the total column sum of haplotype
$c$ and $H_S=\sum_{s=1}^S 1/s$. 
Finally, the model is completed with a prior on 
$\bw_s = (w_{s0},\ldots,w_{sC})$, $s=1,\ldots,S$. We assume independent
Dirichlet priors, independent across SNVs,
$\bw_s \sim \Dir(a_0,\ldots,a_C)$, using a common value
$a_1=\ldots=a_C=a$ and distinct $a_0$.

In summary, the hierarchical model factors as 
\begin{equation}
   p(\bZ, \bw, \bn \mid \bN) = 
   \underbrace{p(\bZ)}_{\eqref{eq:IBP1}}
   p(\bw\mid \bZ)
   \underbrace{p(\bn\mid \bp,\bN)}_{\eqref{eq:like}}.
\label{eq:model}
\end{equation}
Recall that $p_{st}$ is specified in
\eqref{eq:model_p} as a deterministic function of $\bZ$ and $\bw$,
that is $p_{st}= p_{st}(\bw,\bZ)$. 
The joint posterior 
\begin{equation}
  p(\bZ, \bw \mid \bN,\bn), 
\label{eq:post}
\end{equation}
and thus 
the desired inference on TH are well defined by \eqref{eq:model}.
However, practically useful inference requires summaries,
such as the MAP (maximum a posteriori) estimate or efficient posterior
simulation from \eqref{eq:model}, which could be used to compute
(most) posterior summaries.
Unfortunately,  both, MAP estimation and posterior simulation are
difficult to implement here.
We therefore extend 
the approach proposed in \cite{broderick2012mad}, 
who define small-variance asymptotics for inference under an IBP
prior using a standard mixture of Gaussian sampling model. 
Below, in Section \ref{sec:MAD}, we develop a similar approach for the binomial model
\eqref{eq:like}, and it applies to any other exponential family
sampling models.

\subsection{Prior Model for Subclones}
Model \eqref{eq:model} characterizes TH by inference on latent
haplotypes. More than $C=2$ haplotypes in one tumor sample
imply the existence of subclones since humans are diploids.
The sequence matrix $\bZ$ and sample proportions $\bw$ for haplotypes
precisely characterize the genetic contents in a potentially
heterogeneous tumor sample. We will use it for most of the upcoming
inference.
However, haplotype inference does not yet characterize subclonal
architecture. A subclone is uniquely defined by a pair of
haplotypes. 
A simple model extension allows inference for subclones, if desired.
As an alternative and to highlight the modeling strategies we briefly
discuss such an extension. 

We introduce a latent trinary $(S \times C)$ matrix
$\tbZ$, with columns now characterizing subclones (rather than
individual haplotypes).
We use three values $\zt_{sc} \in \{0, 1,2\}$ to represent the subclonal
true VAFs at each locus, with $\zt_{sc}=0$ indicating homozygosity and no variant on
both alleles, and $\zt_{sc}=1$ and $2$ indicating a heterozygous variant and a
homozygous variant, respectively. 
The true VAF at each SNV is equivalent to the subclonal genotype
and also known as B-allele  frequency. 
We use the unconventional term ``true VAF" to be consistent
of our previous discussion. 

We define a prior model $p(\tbZ)$ as a variation of the IBP prior.
Let $m_{c1}=\sum_s I(\zt_{sc}=1)$ and  $m_{c}=\sum_s I(\zt_{sc}>0)$, where $I(\cdot)$ is the
indicator function and define
\begin{equation}
  p(\tilde{\bZ} \mid \gamma) =
  \frac{\gamma^C\, e^{-\gamma H_S}}
       {C!}
\;
  \prod_{c=1}^C
  \frac{(S-m_c)!(m_c-1)!}
       {S!}\pi_c^{m_{c1}}(1-\pi_c)^{m_c-m_{c1}}.
       \label{eq:IBP2}
\end{equation}
The construction can be easily explained. Starting with an IBP prior $p(\bZ)$
for a latent binary matrix $\bZ$, we interpret $z_{sc}=1$ 
as an indicator for variant $s$ appearing in subclone $c$ (homo-
{\em or} heterozygously), that is 
$\{z_{sc}=1\} = \{\zt_{sc} \in \{1,2\} \}$.   
For each $z_{sc}=1$ we flip a coin. With probability $\pi_c$ we record
$\zt_{sc}=1$ and with probability $(1-\pi_c)$ we record $\zt_{sc}=2$.
And we copy $z_{sc}=0$ as $\zt_{sc}=0$. 
 
Finally, we assume that each sample consists of proportions $w_{tc}$
of the subclones, $c=1, \dots, C$, and represent $p_{st}$ as 
$$
  p_{st} = w_{t0}p_0 + \frac{1}{2}\sum_{c=1}^{C} w_{tc} \zt_{sc}. 
$$
Here, the decomposition of $p_{st}$ includes again an additional term
for a background subclone.
Lastly, we complete the model with a uniform prior on $\pi_c$, $c=1,
\dots, C$ and denote $\bpi=(\pi_1, \dots, \pi_C)$.  The hierarchical
model for estimating subclones factors as  
\begin{equation}
   p(\tbZ, \bw, \bpi, \bn \mid \bN) = 
   \underbrace{p(\tbZ, \bpi)}_{\eqref{eq:IBP2}}
   p(\bw\mid \tbZ)\, 
   \underbrace{p(\bn\mid \bp,\bN)}_{\eqref{eq:like}}.
\label{eq:model2}
\end{equation}
We will use model \eqref{eq:model2} for alternative inference on
subclones, but will focus on inference for haplotypes under model
\eqref{eq:model}. That is, we characterize TH by decomposing observed
VAF's in terms of latent haplotypes.

\section{A MAD Bayes Algorithm for TH}
\label{sec:MAD}

\subsection{Bregman Divergence and the Scaled Binomial Distribution}

We define small-variance asymptotics for any exponential family
sampling model, including in particular the binomial sampling model
\eqref{eq:like}.
The idea is to first rewrite the exponential family model in the
canonical form as a function
of a generalized distance between the random variable and the mean
vector. We use Bregman divergence to do this. In the canonical form it is then
possible to define a natural scale parameter which becomes the target
of the desired asymptotic limit.
Finally, we will
recognize the log posterior as approximately equal to a K-means type
criterion. The latter will allow fast and efficient evaluation of
the MAP. Repeat computations with different starting values finds a
set of local modes, which are used to summarize the posterior
distribution. 
The range of local modes gives some information about the
effective support of the posterior distribution.  
We start with a definition of Bregman divergence.

\begin{definition}
Let $\phi: \mathcal{S}\to \mathbb{R}$ be a differentiable, strictly
convex function defined on a convex set $\mathcal{S}\subseteq
\mathbb{R}^n$. The Bregman divergence  \citep{bregman1967relaxation}
for any points $\bx, \by \in \mathbb{R}^n$ is defined as  
$$
   d_{\phi}(\bx,\by)=
   \phi(\bx)-\phi(\by)-\langle \bx-\by, \nabla\phi(\by) \rangle,
$$
where $\langle \cdot, \cdot \rangle$ represents the inner product and
$\nabla\phi(\by)$ is the gradient vector of $\phi$.
\end{definition}
In words, $d_\phi$ is defined as the increment $\{\phi(\bx)-\phi(\by)\}$
beyond a linear approximation with the tangent in $\by$. 
The Bregman divergence leads to a large number of useful divergences
as special cases,  such
as squared loss distance, KL-divergence, logistic loss, etc. For
instance, if $\phi(\bx)=\langle \bx, \bx \rangle$, 
$d_{\phi}(\bx,
\by)=||\bx-\by||^2$ is the squared Euclidean distance. 

\cite{banerjee2005clustering} show that there exists a unique Bregman
divergence corresponding to every regular exponential family including
binomial distribution. Specifically, defining the natural
parameter $\eta=\log(\frac{p}{1-p})$, we rewrite the probability mass
function of $n \sim \Binom(N, p)$ in the canonical form, given by 
\begin{equation}
  p(n\mid \eta, \psi) 
        =\exp\{n\eta-\psi(\eta)-h_1(n)\}, 
\label{eq:canon}
\end{equation}
where $\psi(\eta)=N\log(1+e^{\eta})$ and $h_1(n)=-\log{N \choose
n}$. Under \eqref{eq:canon}  the mean and variance
of $n$ can be written as a function of $\eta$, given by 
\begin{equation}
  \mu=\nabla\psi(\eta)=N\frac{e^{\eta}}{1+e^{\eta}} \quad \mbox{and}
  \quad 
  \sig^2=\nabla^2\psi(\eta)=N\frac{e^{\eta}}{(1+e^{\eta})^2}.
\label{eq:muV}
\end{equation}

We introduce a rescaled version of the likelihood by a power
transformation of the kernel of \eqref{eq:canon}, that is by scaling
the first two terms in the exponent, replacing $\eta$ by $\teta =
\beta\, \eta$ and $\psi(\eta)$ by 
  $\tpsi(\teta)  = \beta \psi(\teta/\beta)$.  
Let $\tp(n \mid \teta, \tpsi)$ denote the power transformed model.
A quick check of \eqref{eq:muV} shows that the mean
remains unchanged,   
$\tmu = \nabla\tpsi(\teta) = \nabla\psi(\eta) = \mu$, and
the variance gets scaled,
$\tsig^2 = \nabla^2 \tpsi(\teta) = \beta \nabla^2 \psi(\teta/\beta) =
\sig^2/\beta$.  
That is, $\tp(\cdot)$ is a rescaled, tightened version of $p(\cdot)$.
The important feature of this scaled binomial model is that
$\tsig^2 \to 0$ as $\beta \to \infty$, while
$\tmu$ remains unchanged. 

The rescaled model can be elegantly interpreted when we rewrite 
\eqref{eq:canon} as a function of Bregman divergence for suitably
chosen $\phi(n,\mu)$. The rescaled version arises when replacing
$d_\phi$ by  $\beta\, d_\phi(n,\mu)$. 
Let
$$
  \phi(n)=n\log(\frac{n}{N})+(N-n)\log(\frac{N-n}{N}).
$$  
The Bregman divergence is 
$d_{\phi}(n, \mu)=\phi(n)-\phi(\mu)-(n-\mu)\nabla\phi(\mu)$, with which the
binomial distribution can be expressed as 
\begin{equation}
  p(n\mid \eta, \psi)=p(n\mid \mu) = \exp\{-d_{\phi}(n, \mu)\}f_{\phi}(n),
\label{eq:exp}
\end{equation}
where $f_{\phi}(n)=\exp\{\phi(n)-h_1(n)\}$. The derivation of
(\ref{eq:exp}) is shown in the Supplementary Material A. 
%
Denoting $\tphi=\beta \phi$, we can write the Bregman divergence
representation for the scaled binomial as 
$$ 
\tp(n\mid\teta, \tilde{\psi}) \equiv \tp(n\mid \tmu)  = \exp\{-d_{\tphi}(n,
\tmu)\}f_{\tphi}(n)
      =\exp\{-\beta d_{\phi}(n, \mu)\}f_{\beta\phi}(n). 
$$ 
For any exponential family model we can write its canonical form and
construct the corresponding Bregman divergence representations. Thus
the same rescaled version can be defined for any exponential
family model.

\subsection{MAP Asymptotics for Feature Allocations}
We use the scaled binomial distribution  
to develop small-variance asymptotics to the hierarchical model
(\ref{eq:model}). 
A similar derivation of small-variance asymptotic to (\ref{eq:model2})
is shown in the Supplementary Material E. 
Let $\tp_\beta(\cdot)$  generically denote distributions under the scaled
model. The joint posterior is  
$$
   L(\bZ,\bw) \equiv \tp_\beta(\bZ, \bw \mid  \bn, \bN)  \propto
     p(\bZ)\, p(\bw\mid \bZ)\, \tp_\beta(\bn\mid \bN,  \bp);
$$
based on \eqref{eq:canon}, the scaled binomial likelihood is given
by 
\begin{multline}
  \tp_\beta(\bn \mid \bN,  \bp)  = 
  \prod_{t=1}^T\prod_{s=1}^S 
  \exp\left\{
    -\beta \left[ n_{st}\log(\frac{n_{st}}{N_{st}p_{st}}) +
      (N_{st}-n_{st})\log(\frac{N_{st}-n_{st}}{N_{st}-N_{st}p_{st}})\right]\right\} \\
  \times \exp\left\{
     \beta n_{st}\left[\log(\frac{n_{st}}       {N_{st}})+(N_{st}-n_{st})
                       \log(\frac{N_{st}-n_{st}}{N_{st}})
                  \right] 
     - h_1(n_{st})\right\}, \nonumber
\end{multline}
where  
$p_{st} = w_{t0}p_0 + \sum_{c=1}^{C} w_{tc} z_{sc}$, as
before. Finding the joint MAP of $\bZ$ and $\bw$ 
is equivalent to finding the values of $\bZ$ and $\bw$ that minimize
$-\log L(\bZ, \bw)$. 
We avoid overfitting $\bZ$ with an inflated number of features
by moving the prior
towards a smaller numbers of features as we increase $\beta$.
This is achieved by varying $\gamma=\exp(-\beta \lambda^2)$ with 
increasing $\beta$, that is $\gamma \to 0$ as $\beta \to \infty$.
Here $\lambda^2>0$ is a constant tuning parameter.
We show that 
\begin{eqnarray}
-\frac{1}{\beta}\log L &\sim& \sum_{s=1}^S\sum_{t=1}^T
 \left\{-n_{st}\log(p_{st})-(N_{st}-n_{st})\log(1-p_{st})\right\}
 +C\lambda^2 \equiv Q(\bp),
\label{eq:objective}
\end{eqnarray}
where $C$ is the random number of columns of $\bZ$, and 
$u(\beta)\sim v(\beta)$ indicates asymptotic equivalence, i.e.,
$u(\beta)/v(\beta) \to 1$ as $\beta\to \infty$.
The double sum originates from the scaled binomial likelihood and the
penalty term arises as the limit of the log IBP prior.
The derivation is shown in the Supplementary Material B. 
We denote the right hand side of \eqref{eq:objective} as $Q(\bp)$.
We refer to $Q(\bp)$ as the FL (feature leaning)-means objective
function,   keeping in mind that 
$p_{st} = w_{t0}p_0 + \sum_{c=1}^{C} w_{tc} z_{sc}$ is a function of
$\bZ$ and $\bw$.  
The first term in the objective functions is a K-means style
criterion for the binary matrix $\bZ$ and $\bw$ when the number of
features is fixed. The second term acts a penalty for the number of
selected features.  The tuning parameter $\lambda^2$ calibrates the
penalty.  
We propose a specific calibration scheme for the application 
  to inference on  
tumor heterogeneity. See Section \ref{sec:lam} for details. A local MAP that maximizes
the joint posterior $L(\bZ, \bw)$
is asymptotically equivalent to
 \begin{equation} 
 \mathrm{argmin}_{(C, \bZ, \bw)}Q({\bp}).
   \label{eq:aim}
 \end{equation}
 Keep in mind that $p_{st}(\bZ,\bw)$ is a function of
$\bZ$ and $\bw$. Objective function $Q(\bp)$ is similar to a
penalized likelihood function, with $\lambda^2$ mimicking the tuning
parameter for sparsity. The similarity between $Q({\bp})$ and a
penalized likelihood objective function is encouraging, highlighting
the connection between an approximated Bayesian computational approach
based on coherent models and regularized frequentist inference. 

\subsection{FL-means Algorithm}
We develop the FL-means algorithm to solve the optimization problem in
(\ref{eq:aim}) and prove convergence.
Since all entries of $\bZ$ are binary and $w_{tc} \in \mathbb{R}^+$
subject to $\sum_{c=0}^Cw_{tc}=1$,  (\ref{eq:aim}) is a mixed
integer optimization problem. Mixed integer linear
programming (MILP) is NP-hard to 
solve \citep{karlof2005integer}. The objective function is a
non-linear function of $\bZ$ and $\bw$, indicating that even
sophisticated MILP solvers unlikely benefit this case. Rather than
solving the optimization problem (\ref{eq:aim}) as a generic MILP, we
construct a coordinate transformation, allowing us to solve this
problem as a constrained optimization problem.  

Denote $\Delta_{Ct}=\{(w_{t0}, \dots, w_{tC})^T\in
\mathbb{R}^{C+1}\mid \sum_{c=0}^Cw_{tc}=1, \mathrm{and} \ w_{tc}\geq
0 \ \mathrm{for \ all} \ c\}.$  Suppose $\bZ$ is fixed.
 Let $\bw_{t,-0}=(w_{t1},\ldots,w_{tC})$ and let \ech
$H(\bZ)
= \{p_0 w_{t0} + \bZ \bw_{t,-0} \mid \bw_t\in \Delta_{Ct}, t=1,
\dots, T\}$ denote the set 
of convex combinations of the column vectors in $\bZ$
(adding the term for the background  
in \eqref{eq:model_p}). 
Then \eqref{eq:aim} reduces to finding 
$$
   \mathrm{argmin}_{\bp\in H(\bZ)}Q(\bp).
$$
It can be shown that the objective function
$Q(\bp)$ is separable convex (see Supplementary Material C), and this problem can
be solved using standard convex optimization methods. However, the
next optimization with respect to 
$\bZ$ given fixed $\bw$ is no longer convex. We
use a brute-force approach to solve this problem by enumerating all
possible $\bZ$. 
We propose the following algorithm for haplotype modeling. It can be easily adapted to subclone modeling.

\paragraph*{FL-means Algorithm}
\begin{itemize}
\item Set $C=1$. Initialize $\bZ$ as an $S\times C$ matrix by
  setting $z_{s1}=1$ with probability 0.5 for $s=1, \dots, S$. Initialize
  $\bw$ as a $T\times (C+1)$ matrix with 
  $\bw_t=(w_{t0}, w_{t1}, \dots, w_{tC}) \sim \Dir(1, 1, \dots, 1)$ for $t=1, \dots, T$.
\item Iterate over the following steps until no changes are made.
  \begin{enumerate}
  \item For $s=1, \dots, S$, minimize $Q$ with respect to 
    $\bz_s=(z_{s0}, z_{s1},\dots, z_{sC})$, fixing $\bw_t$ and $C$ at
    the 
    currently imputed values. 
  \item For $t=1, \dots, T$, minimize $Q$ with respect to 
    $\bw_t=(w_{t0}, w_{t1}, \dots, w_{tC})$ with constraint
    $\sum_{c=0}^Cw_{tc}=1$, fixing $\bZ$ and $C$ at the 
    currently imputed values. 
  \item Let $\bZ'$ equal $\bZ$ but with one new feature (labeled
    $C+1$) containing only one randomly selected SNV  index $s$. Set
    $\bw'$ that minimizes the objective given $\bZ'$. If the triplet
    $(C+1, \bZ', \bw')$ lowers the objective $Q$ from the
    triplet $(C, \bZ, \bw)$, replace the latter with the
    former. 
  \end{enumerate}
\end{itemize}

\begin{theorem}\label{tm:property}
  The FL-means algorithm converges in a finite number of iterations
  to a local minimum of the FL-means objective function $Q$. 
\end{theorem}
See Supplementary Material D for a proof. 
Theorem \ref{tm:property}  guarantees convergence, but does not guarantee convergence
to the global optimum.  From a data analysis perspective the presence
of local optima is a feature.  It can be exploited to learn about the
sensitivity  to initial conditions and degree of multi-modality by using multiple random
initializations. We will demonstrate this use in Sections
\ref{sec:simu} and \ref{sec:analysis}.

\subsection{Evaluating Posterior Uncertainty}
\label{sec:pbar}
The FL-means algorithm implements computationally efficient evaluation of
an MAP estimate $(\Chat,\Zhat,\what)$ for the unknown model
parameters. However, a major limitation of any MAP estimate is the lack
of uncertainty assessment. We therefore supplement the MAP report
with a summary of posterior uncertainty based on the conditional posterior
distribution $p(\bZ \mid \Chat, \bn, \bN)$. 
In particular, we report
\begin{equation}
  \pbar_{sc} \equiv p(z_{sc}=\zhat_{sc} \mid \Chat, \bn, \bN).
\label{eq:uncertainty}
\end{equation}
Evaluation of $\pbar_{sc}$ is easily implemented by posterior MCMC
simulation.
Conditional on a fixed number of columns, $\Chat$, we can implement posterior
simulation under \eqref{eq:post} (for fixed $\Chat$) using Gibbs sampling transition
probabilities to update $z_{sc}$ and Metropolis-Hastings transition
probabilities to update $w_{tc}$. 
We initialize the MCMC chain with the MAP estimates
$\bZ=\Zhat$ and $\bw=\what$.
See Supplementary Material F for the details of the MCMC.
The MCMC algorithm is implemented with 1,000 iterations and used to
evaluate $\pbar_{sc}$. In the following examples we report $\pbar_{sc}$
to characterize uncertainty of the reported inference for TH.

\section{Simulation Studies}
\label{sec:simu}
\subsection{Haplotype-Based Simulation }
\label{sec:simu1}
We carried out simulation studies to evaluate the proposed FL-means
algorithm for haplotype inference, i.e., using model \eqref{eq:model}.   
\begin{figure}[hbtp]
  \centering{
    \begin{tabular}{ccc}
      \includegraphics[width=.32\textwidth]{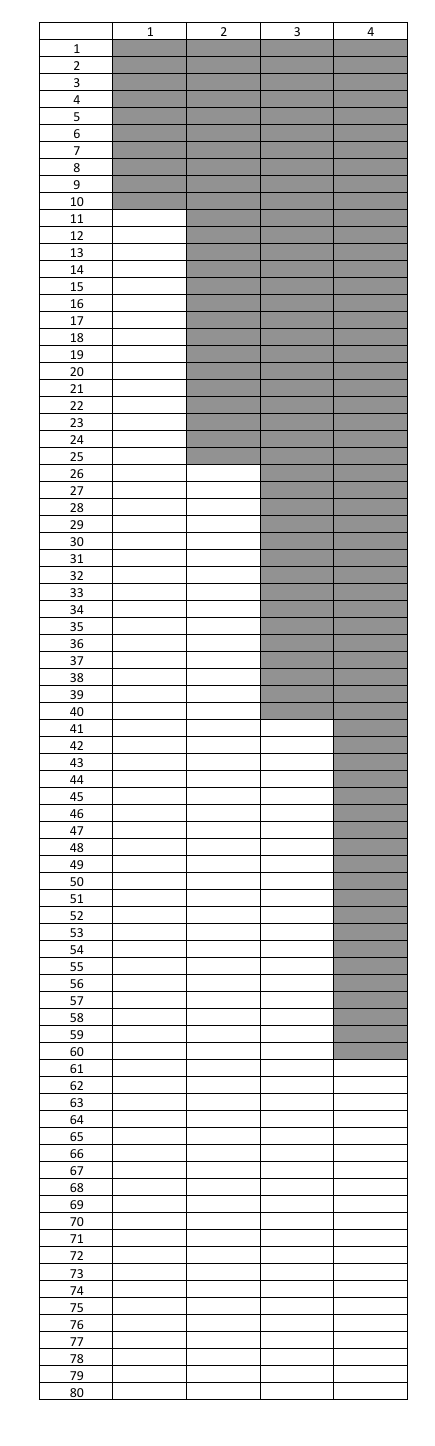}&
      \includegraphics[width=.32\textwidth]{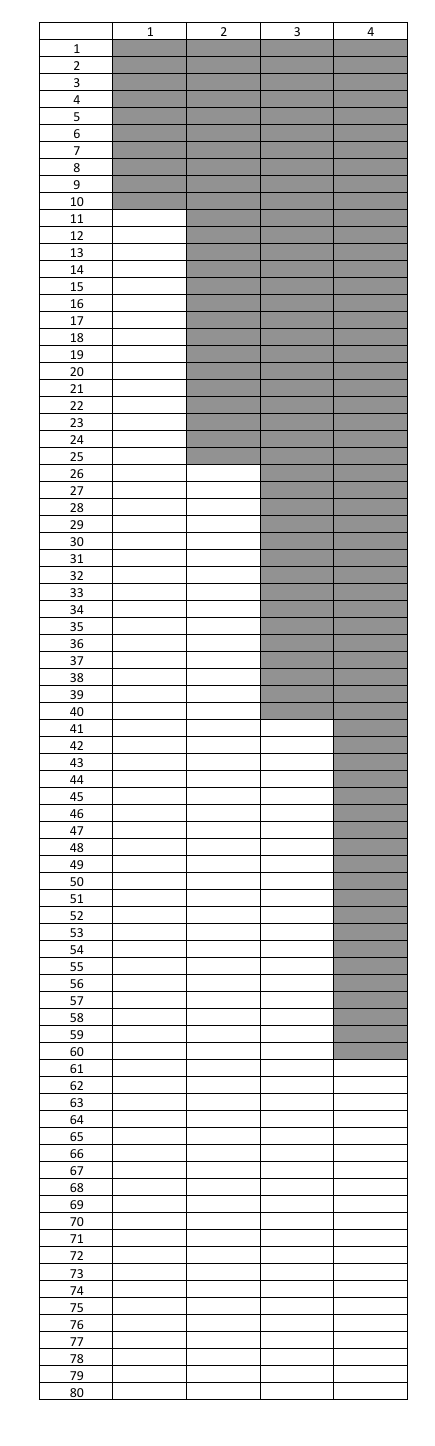}&
      \includegraphics[width=.32\textwidth]{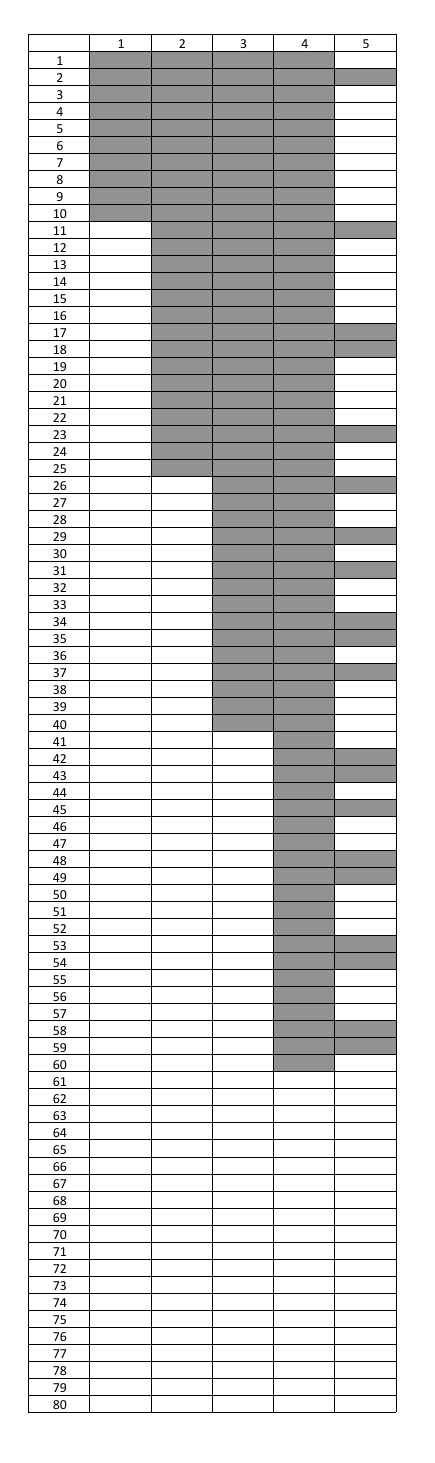}\\
      (a) Simulation Truth & (b) $\Zhat$ (under $\lambda^2=10, 8$) & 
      (c) $\Zhat$ ($\lambda^2=6$)\\
    \end{tabular}
  }
\caption{A simulation example. The plots show the feature allocation
  matrix $\bZ$,  with shaded area indicating $z_{sc}=1$, i.e., variant genotype. 
  Rows are the SNVs and columns are the inferred haplotypes.  
  Panel (a) shows the simulation truth ${\bZ}^{\TRUE}$. Panel (b) displays 
  the estimated feature allocation matrix $\Zhat$  for $\lambda^2=8$
  and $10$.
  The estimate perfectly recovers the simulation truth. Panel (c)
  shows $\Zhat$ for $\lambda^2=6$. The first four haplotypes match the simulation
  truth.} 
 \label{fig:simu}
 \end{figure}
We generated a data matrix with $S=80$ SNVs and $T=25$ samples. 
The simulation truth included 
$C^\TRUE=4$ latent haplotypes, plus a background haplotype  that included all SNVs. The latent binary matrix
${\bZ}^{\TRUE}$ was generated as follows: haplotype 1 was characterized
by the presence of the first 10 SNVs, haplotype 2 by the first 25,
haplotype 3 by the first 40 and haplotype 4 by the first 60. In other
words, SNVs 1-10 occurred in all four haplotypes, SNVs 10-25 in haplotypes 2-4, SNVs 25-40 in haplotypes 3-4, SNVs 41-60 in haplotype 4
only,  and SNVs 61-80 in none of the haplotypes. 
Figure \ref{fig:simu}(a) shows the simulation truth $\bZ^\TRUE$. Let
$\pi=(1,5,6,10)$ and $\pi_{p}$ be a random permutation of $\pi$, we
generated true $\bw_t^{\TRUE}\sim \mathrm{Dirichlet}(0.2, \pi_p)$, $t=1,
\dots, T$.
Let $p^o_0=0.01$ and $N_{st}=50$ for all $s$ and $t$; we
generated $n_{st}\sim \mathrm{Bin}(N_{st}, p_{st})$, where
$p_{st}= p_0^o w^{\TRUE}_{t0}+\sum_{c=1}^{C^\TRUE}w_{tc}^{\TRUE}z_{sc}^{\TRUE}$. 
We then ran the FL-means algorithm repeatedly with 1,000 random
initializations to obtain a set of local minima of
$Q$ as an approximate representation of posterior
uncertainty.  
Each
run of the FL-means algorithm only took 1 minute.

\begin{figure}[bthp]
\centering
\includegraphics[width=.6\textwidth]{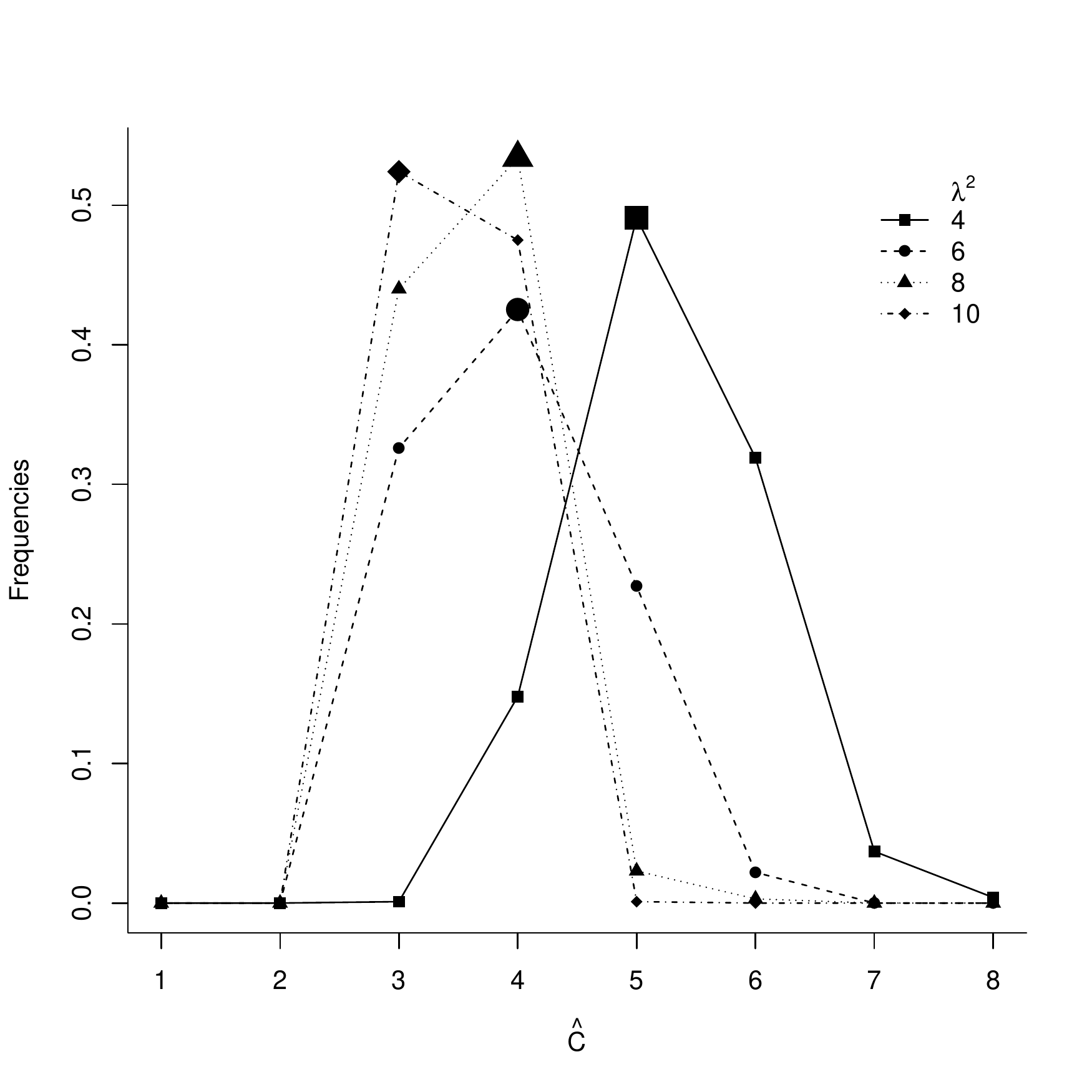}
\caption{The frequencies of estimated $\Chat$  in 1000
  simulations with different initializations. The mode is highlighted
  with extra large plotting symbols.}
\label{fig:hist}
\end{figure}

For different $\lambda^2$ values, we report point estimates (Figure \ref{fig:simu} (b) and (c)) for $\bZ$ that were obtained
by minimizing the objective function $Q$.  The point
estimate is the estimate $(\Chat, \Zhat, \what)$ that
minimizes $Q(\bp)$ in 1,000 random initializations. 
 Figure \ref{fig:hist} shows the frequencies  of the estimated
number $\Chat$ of features under different $\lambda^2$ values (the modes
are highlighted by extra large plotting symbols).
The plot shows the distribution of local minima over
repeated runs of the algorithm with different initializations, all
with the same simulated data.

An important consideration in the implementation of the FL-means
algorithm is the choice of the 
penalty parameter $\lambda^2$.
As a sensitivity analysis we ran the FL-means algorithm
with different $\lambda^2$ values: $\lambda^2=2, 4, 6, 8, 10, 20,
200$ and 500. Figure S1 (see Supplementary Material G) shows the estimated number 
$\Chat$ of features and the realized minimum value of the objective function $Q$
versus $\lambda^2$. 
We observe that the $\Chat$ decreases and the objective value
increases as $\lambda^2$ increases. Summarizing Figures \ref{fig:simu}
and S1, we find that under $\lambda^2=8, 10$, both the
estimated $\Chat$ and the 
estimated $\Zhat$ perfectly reconstruct the simulation truth
$\bZ^\TRUE$. 
Under $\lambda^2=6$, we find $\Chat=5$ and the estimated $\Zhat$
includes true haplotypes (columns 1-4) as well as an additional 
spurious haplotype that includes some of the SNVs. 
We add a report of posterior uncertainties $\pbar_{sc}$, as
defined in \eqref{eq:uncertainty}. In this particular case we find
$\pbar_{sc}=1$ for all $s, c$. The estimate $\Zhat$ recovers the
simulation truth, and there is little posterior uncertainty about it.
As expected, a
smaller $\lambda^2$ leads to a larger $\Chat$ value. 
 In summary, the inference summaries are sensitive with respect to
the choice of $\lambda^2$. A good choice is critical. Below we suggest
one reasonable ad-hoc algorithm for the choice of $\lambda^2$. 

For comparisons, we applied a recently published method PyClone
\citep{roth2014pyclone} to the same simulated data.  PyClone
identifies candidate subclones as individual clusters by 
partitioning  the SNVs into sets of similar VAFs.  However, PyClone does not
allow overlapping sets, that is, SNVs can not be shared across 
subclones, and is thus not fully in line with clonal evolution theory.  
Inference is still meaningful as a characterization of heterogeneity,
but should be interpreted with care. 
To
implement PyClone, we assumed that the copy number at SNVs was
known. PyClone identified four clusters:  cluster 1 consists of
SNVs 1-25,  cluster 2 SNVs 26-40, cluster 3 SNVs 41-60, and cluster 4
SNVs 61-80. The estimated cluster 1 includes the SNVs that appear
in all the true haplotypes under the simulation truth; cluster 2
includes the SNVs from true haplotypes 3 and 4; cluster 3 includes the
SNVs from true haplotypes 4 and cluster 4 includes the SNVs from none
of the true haplotypes. Figure S2 plots the estimated mean cellular
prevalence of each cluster across all the 25 samples. To summarize, in
our simulation study PyClone did not recover the true haplotypes,
which can not possibly be captured with non-overlapping clustering.

 \subsection{Subclone-Based Simulation} 
We carried out one more simulation study to
evaluate the proposed FL-algorithm under model (\ref{eq:model2}) for
subclonal inference.
We generated a data matrix with $S=80$ SNVs and
$T=25$ samples. The simulation truth included $C^\TRUE=4$ latent
subclones, plus a background subclone that included all
SNVs. We generated the latent trinary matrix
$\widetilde{{\bZ}}^{\TRUE}$ as follows. We first generated a binary
matrix ${\bZ}^{\TRUE}$ as before, in  Section \ref{sec:simu1}.
Next, if $z^\TRUE_{sc}=1$ then 
set $\zt^\TRUE_{sc}=j$ with probabilities 0.7 (for $j=1$) and 0.3
(for $j=2$).
If  $z^\TRUE_{sc}=0$ then $\zt^\TRUE_{sc}=0$.
Figure S3(a) (Supplementary Material G) shows
the simulation truth $\widetilde{\bZ}^\TRUE$. 
 
Next we fixed $p^o_0=0.01$ and $N_{st}=50$ and generated a simulation truth
$\bw_t^{\TRUE}$ as before, in Section \ref{sec:simu1}.  
Finally, we generated   
$n_{st}\sim \Bin(N_{st}, p_{st})$, where 
$p_{st}= p_0^o w^{\TRUE}_{t0}+\frac{1}{2}\sum_{c=1}^{C^\TRUE}w_{tc}^{\TRUE}\zt_{sc}^{\TRUE}$.
We then ran the FL-means algorithm repeatedly with 1,000 random
initializations to obtain a set of local minima of
$Q$. 
The set of local minima provides the desired summary of the posterior
distribution on the decomposition into subclones.
We fix $\lambda^2$ using the algorithm from Section \ref{sec:lam}.
As a result we estimated $\Chat=4$, that is, we estimated the 
presence of four subclones. Figure S3(b) shows the estimated $\bZt$ under
$\lambda^2=50$. In fact, posterior inference in this case perfectly
recovered the simulation truth.

\subsection{Calibration of $\lambda^2$} 
\label{sec:lam}
Recall that $w_{tc}$ denotes the relative fraction of haplotype $c$ in
sample $t$. Under $\lambda^2=8$, the posterior estimate 
perfectly recovered the simulation truth.
The estimates for $w_{tc}$ ranged 
from 0.01 to 0.53 for $c=1$, 
from 0.007 to 0.57 for $c=2$, 
from $1.8\times 10^{-8}$ to 0.52 for $c=3$ and 
from 0.008 to 0.59 for $c=4$ for $t=1, \dots, T$. 
Posterior inference (correctly) reports that each true haplotype
$c$ constitutes a substantial part of the composition for some subset
of samples.  
Under $\lambda^2=6$,  the first
four estimated features perfectly recover the simulation
truth.
However, for $c=5$,
the estimated $w_{tc}$ ranged from $1.1\times 10^{-9}$ to 0.06.
These very small fractions are biologically meaningless. 
We find similar patterns under $\lambda^2=2$ and
4. Based on these observations, we propose a heuristic to fix the 
tuning parameter $\lambda^2$. We start with a large value of
$\lambda^2$, say, $\lambda^2=50$. While every imputed haplotype $c$
constitutes a substantial fraction in some subset of samples, 
say $w_{tc}>1/\Chat$ for some $t$, we decrease $\lambda^2$ until newly
imputed haplotypes only take small fraction in all samples, say,
$w_{tc}<1/\Chat$. The constant specified here is not arbitrary, instead, it can be chosen based on the biological questions that the researchers aim to address. For example, the haplotype  prevalence below certain threshold is likely to be noise.

\section{Results}
\label{sec:analysis}
\subsection{Intra-Tumor Heterogeneity}
We use deep DNA-sequencing data from an in-house experiment
to study intra-TH that is characterizing heterogeneity of
multiple samples in a single tumor. 
Data include
four surgically dissected tumor samples from the same lung cancer
patient.
We performed whole-exome sequencing and processed the data  using a bioinformatics
pipeline consisting of standard procedures, such as base calling,
read alignment, and variant calling. SNVs with VAFs
close to 0 or 1 do not contribute to the heterogeneity analysis
since these VAF values are expected when samples are homogeneous. We
therefore remove these SNVs from the analysis. 
Details are given in the Supplementary Material H. The final number of
SNVs for  the four intra-tumor samples is $17,160$. With such a large data size, in practice it is
infeasible to run the MCMC sampler proposed in
\cite{juhee2013feature}. 
PyClone is not designed to handle large-scale data sets.  It took more than three days without returning any
result, which makes it practically infeasible for high-throughput data
analysis.

\begin{figure}[h!]
  \begin{center}
    \begin{tabular}{ccc}
      \includegraphics[width=.32\textwidth]{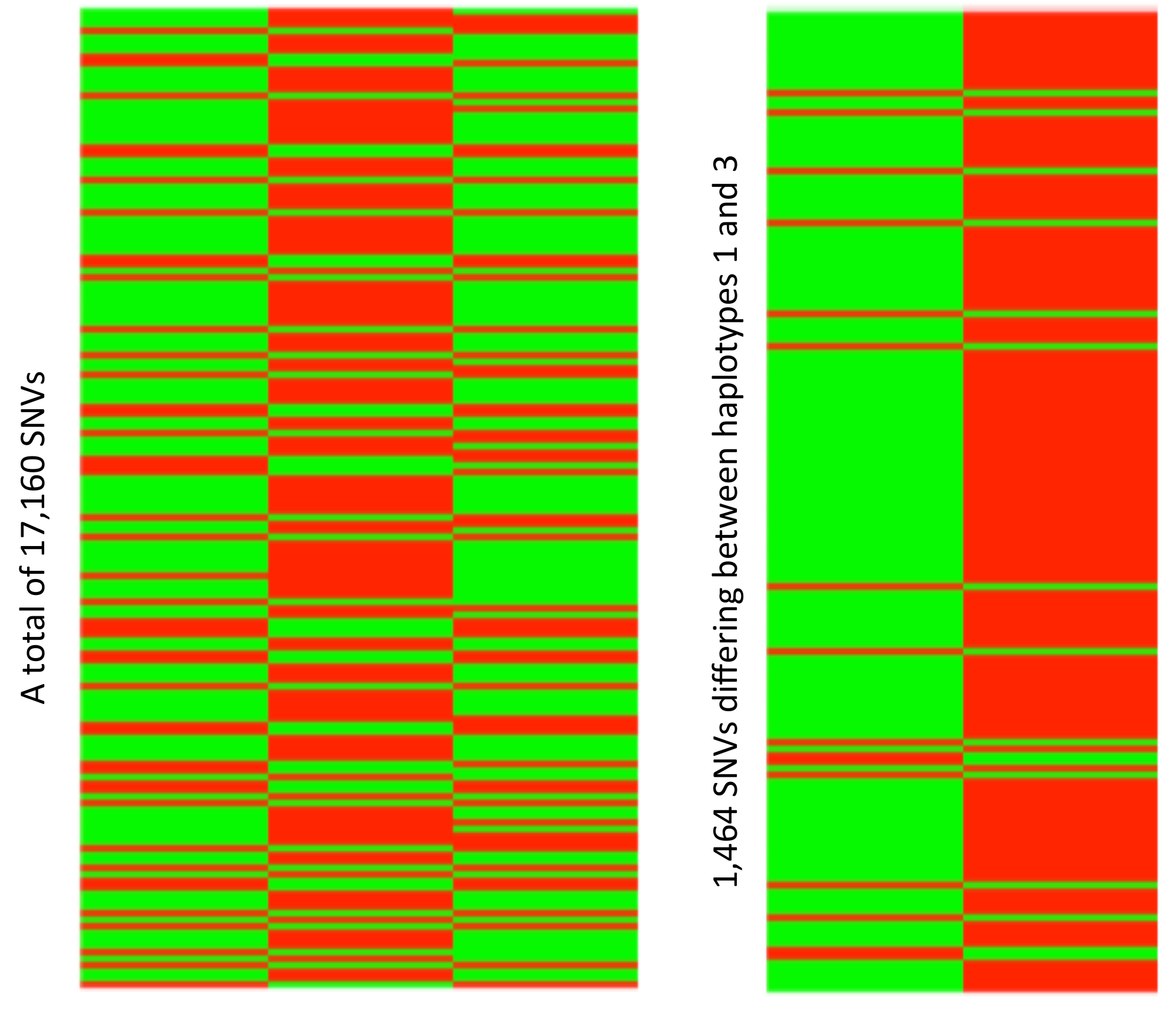}
      &
      \includegraphics[width=.32\textwidth]{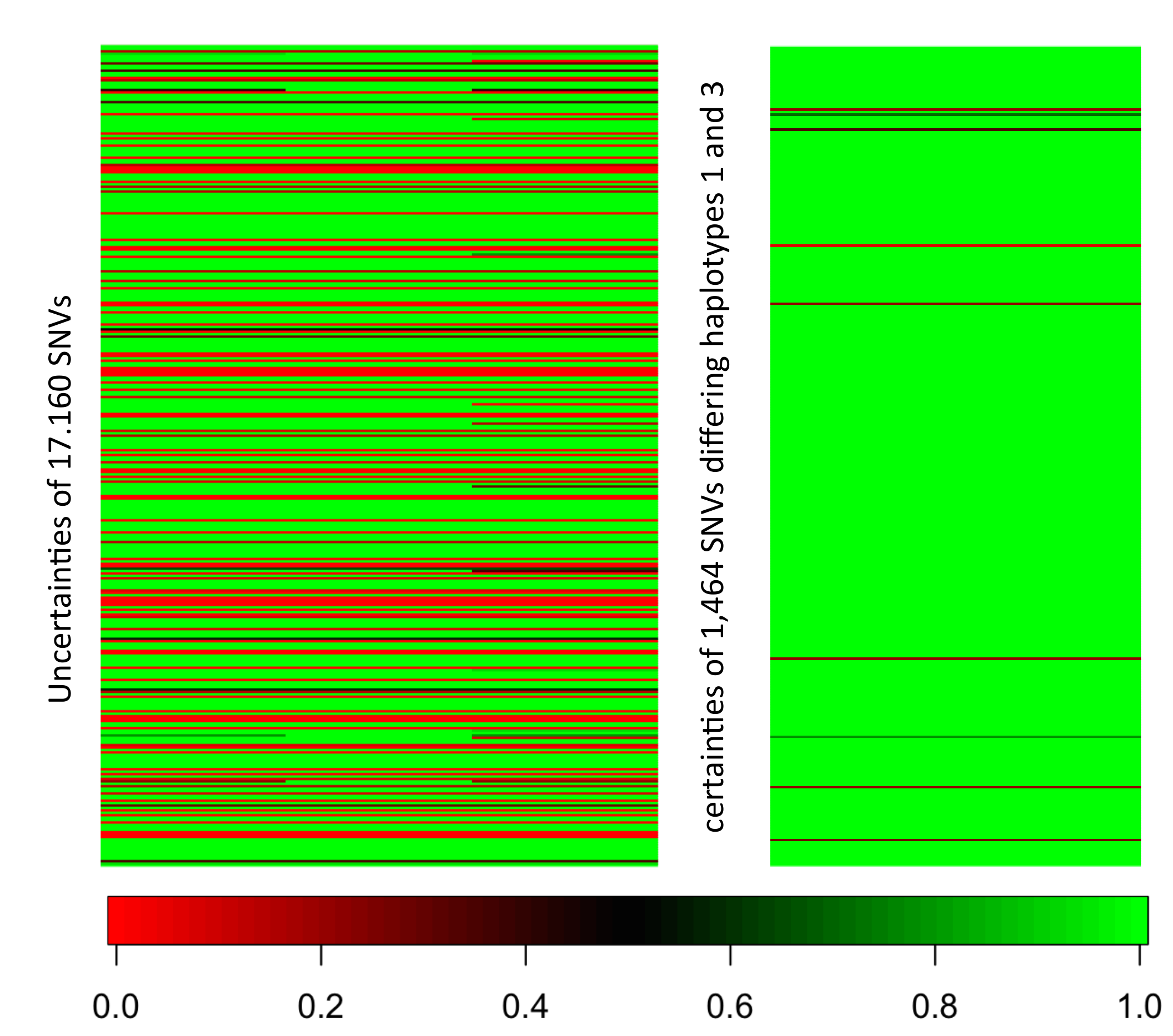}
      &
      \includegraphics[width=.3\textwidth]{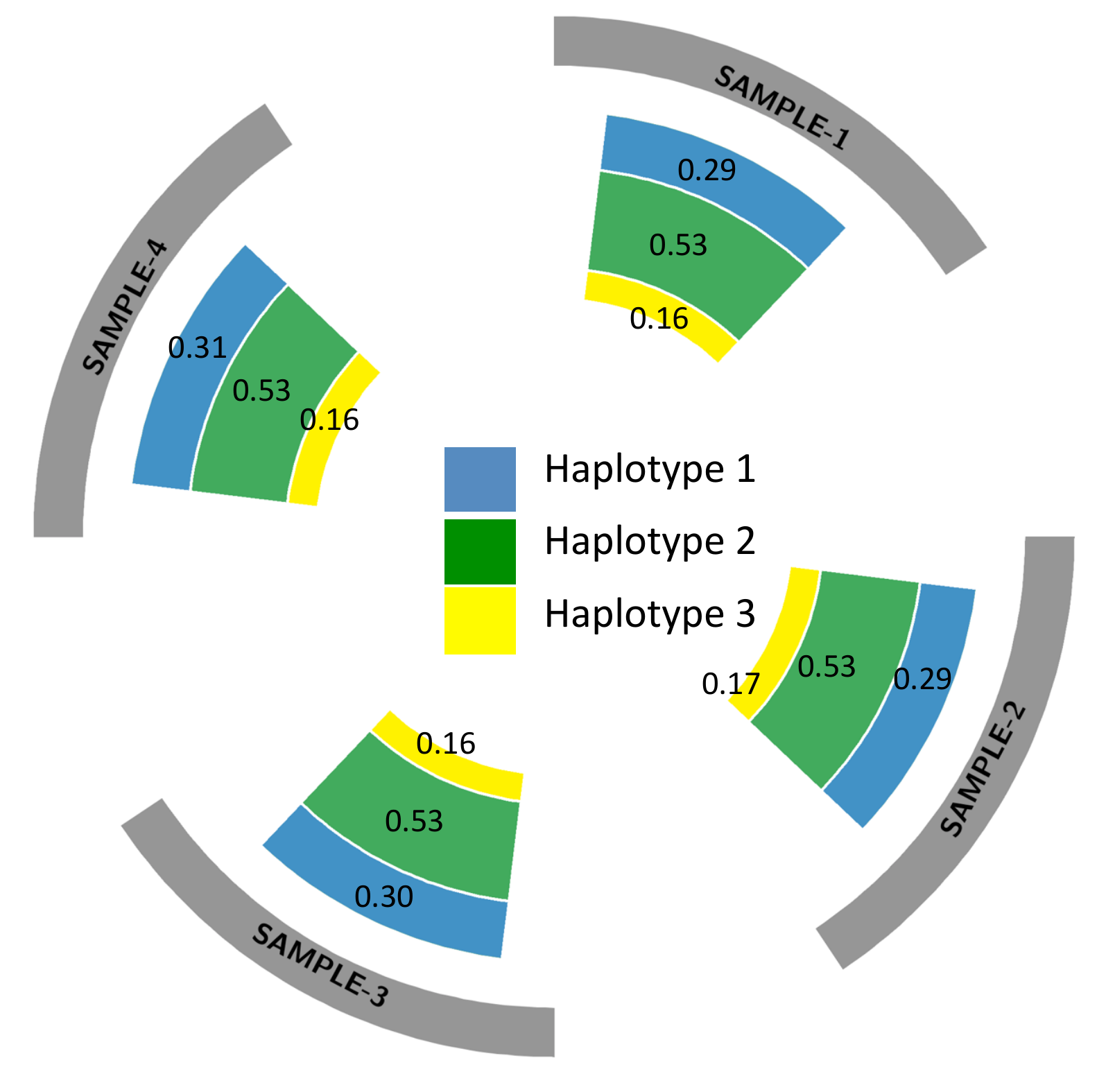} \\
      (a) & (b) & (c) \\
    \end{tabular}
  \end{center}
  \caption{ Summary of intra-TH analysis for four samples from a lung
    tumor with 17,160 SNVs. (a): The first heatmap shows posterior
    estimated haplotype genotypes (columns) of the selected SNVs. Three
    haplotypes are identified. A green/red block indicates a
    mutant/wildtype allele at the SNV in the haplotype. The second heatmap
    shows the differences between haplotypes 1 and 3 in the first
    heatmap. (b): The first heatmap shows the estimated uncertainties of three estimated haplotypes for 17,160 SNVs. The second heatmap
    shows the estimated uncertainties of 1,464 SNVs that differentiate haplotypes 1 and 3 in the first
    heatmap. (c): A circos plot showing the estimated proportion
    $\hat{w}_{tc}$ with $\Chat=3$ haplotypes in each sample. The four
    samples possess similar proportions.}
  \label{fig:data4}
\end{figure}

We first report inference based on haplotypes. Figure
\ref{fig:data4} summarizes the results. 
We fix $\lambda^2$ using the algorithm from Section
\ref{sec:lam}, and find $\Chat=3$ haplotypes. 
In Figure \ref{fig:data4}(a), we observe that haplotypes 1
and 2 have exactly complementary genotypes. 
Next we compare haplotypes 1 and 3. The second heatmap
(still in panel a) plots only the 1,464 SNVs that differ between
$c=1$ and $c=3$, highlighting the differences that are 
difficult to see in the earlier plot over all 17,160 SNVs. 
Three reported haplotypes imply that there are at least two
subclones of tumor cells, one potentially with heterozygous mutations on the 17,160
SNVs, and another with an additional 1,464 somatic mutations. We
annotated the 1,464 mutations and found that a large proportion of
the mutations occur in
known lung cancer biomarker genes. Some of them are {\it de novo}
findings that will be further investigated. 

Next we summarize posterior uncertainty by plotting 
$\pbar_{sc}$, as proposed in \eqref{eq:uncertainty}.
This is shown in Figure \ref{fig:data4}(b).
Green (light grey, $\pbar_{sc}=1$) means no posterior
uncertainty. The plot in panel (b) is arranged exactly as in panel
(a), with rows corresponding to SNVs and columns corresponding to
haplotypes. The left plot shows all SNVs for all three
haplotypes. The right plot zooms in on the subset of 1,464 SNVs that
differ across $c=1$ and $3$ only, similar to panel (a).
There is little posterior uncertainty about the 1,464
SNVs that differentiate haplotype 1 and haplotype 3.
Figure \ref{fig:data4}(c) 
shows the circos plot \citep{Krzywinski:2009} including the estimated
proportion $\hat{w}_{tc}$ 
with $\Chat=3$ haplotypes in each sample.  Interestingly, four tumor
samples possess similar proportions, indicating lack of spatial
heterogeneity across four samples. This is not surprising as the four
samples are taken from tumor regions that are geographically close.

Next we consider a separate analysis using model \eqref{eq:model2}
for inference on subclones. 
Posterior inference estimates $\Chat=3$ subclones, and all four
samples have similar proportions of the three subclones, with two
subclones taking about 40\% and 45\% of the tumor content, and the third subclone
taking about 15\% of the tumor content in all four samples. This agrees with the
reported haplotype proportions in Figure \ref{fig:data4}. However, the
genotypes of subclones are expressed at 0, 1 or 2's, representing
homozygous reference, heterozygous and homozygous variant,
respectively. 
 Inference on subclones does not include inference on
the constituent haplotypes, making it impossible to directly compare
with the earlier analysis.
However, we note that both analyses show that 
the four tumor samples are subclonal, and the subclone
proportions are similar in all four samples. As a final model
checking, 
Figure S4 shows the differences of $(p_{st} - \hat{p}_{st})$,
where $\hat{p}_{st}$ is computed by plugging in the posterior
estimates of $\bw$ and $\bZ$. As the figure shows, both analyses
fit the data well.

\subsection{Inter-Patients Tumor Heterogeneity} 
We analyzed exome-sequencing data for five tumor 
samples from patients with pancreatic
ductal adenocarcinoma (PDAC)  at NorthShore University
HealthSystem 
\citep{juhee2013feature}. Since samples were from different patients, we aimed to infer 
inter-patient TH.  We applied models \eqref{eq:model} and
\eqref{eq:model2} for haplotype- and subclone-based inference,
respectively. In both applications, we focused on a set of SNVs, and assumed that
 the collection of the genotypes at the SNVs could be shared between
 haplotypes in different tumor samples, regardless of the rest of the genome. 


The mean  sequencing depth 
 for the samples was between 60X and 70X.  A total of
approximately 115,000 somatic SNVs were identified across
the five whole exomes using GATK \citep{mckennathe2010}. 
First we focus on a small
number of 118 SNVs  selected with 
the following three criteria: 
(1) exhibit significant coverage in all samples; (2)
occur within genes that are 
annotated to be related to PDAC in the KEGG
pathways database \citep{kanehisa2010kegg}; 
(3) are nonsynonymous, i.e., the mutation changes the amino
acid sequence  that is  coded by the gene.

In summary, the PDAC data recorded the total read counts ($N_{st})$
and variant read counts ($n_{st}$) of $S=118$ SNVs from
$T=5$ tumor samples. Figure S5 (see Supplementary Material G)
 shows the histogram of the observed VAFs, 
$n_{st}/N_{st}$. We ran the proposed FL-means algorithm 
with 1,000 random initializations.  Each run of the FL-means algorithm
took 50 seconds, while an 
MCMC sampler for finite feature allocation model proposed by
\cite{juhee2013feature} took over one hour for each fixed number of
haplotypes. 
Unlike MCMC, iterations of the FL-means algorithm are independent of
each other.
This facilitates parallel computing. For example, 
running all 1000 FL-means repetitions simultaneously the entire
computation finished within one minute (instead of $>$10 hours if
running sequentially). This is a significant speed advantage over MCMC
simulation.
%
After searching for reasonable values of $\lambda^2$ using the
algorithm from Section \ref{sec:lam},
we estimated $\Chat$ to be 5, i.e., five haplotypes.

\begin{figure}
\begin{center}
\begin{tabular}{cc}
\includegraphics[width=.35\textwidth]{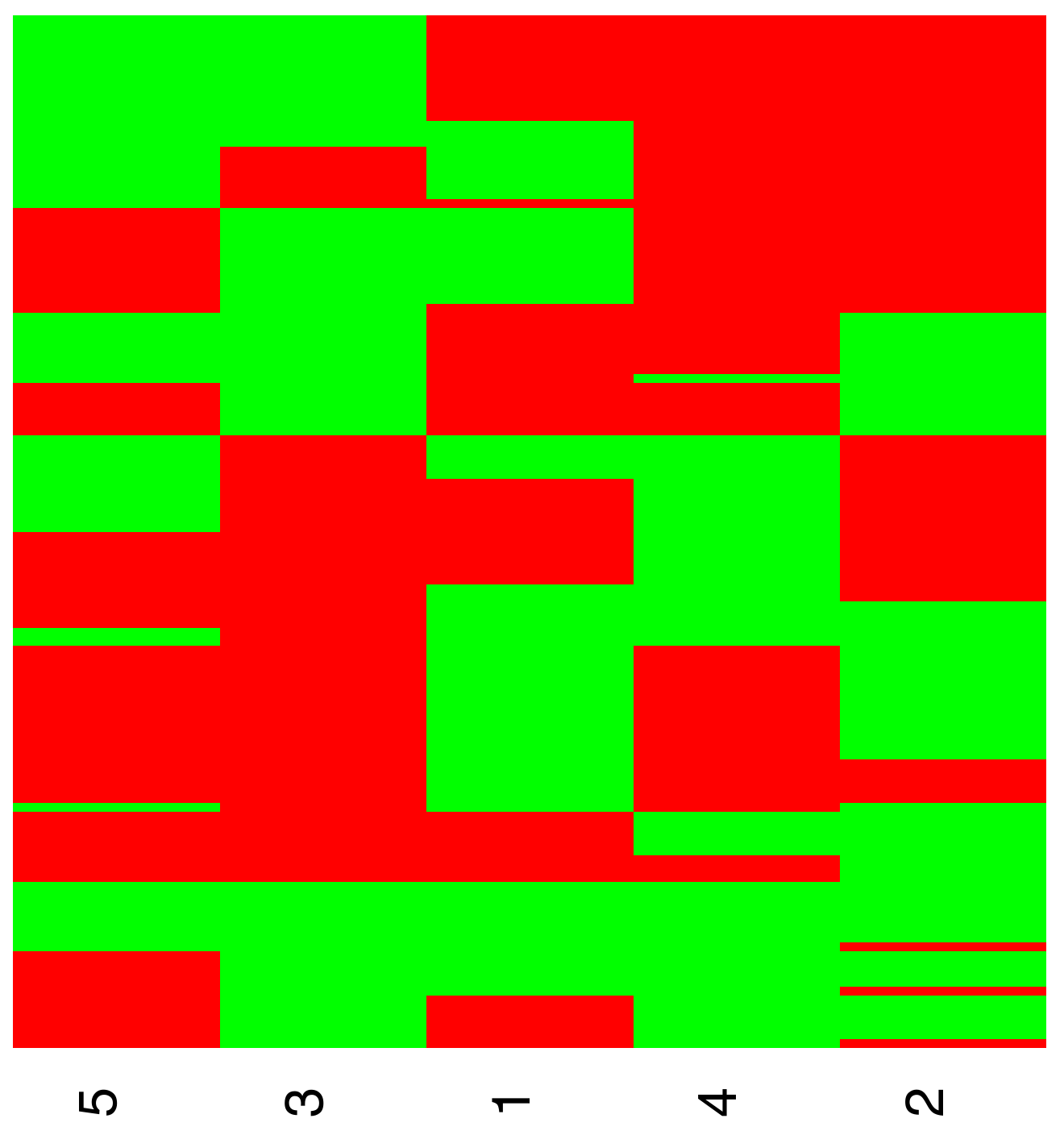}
&
\includegraphics[width=.39\textwidth]{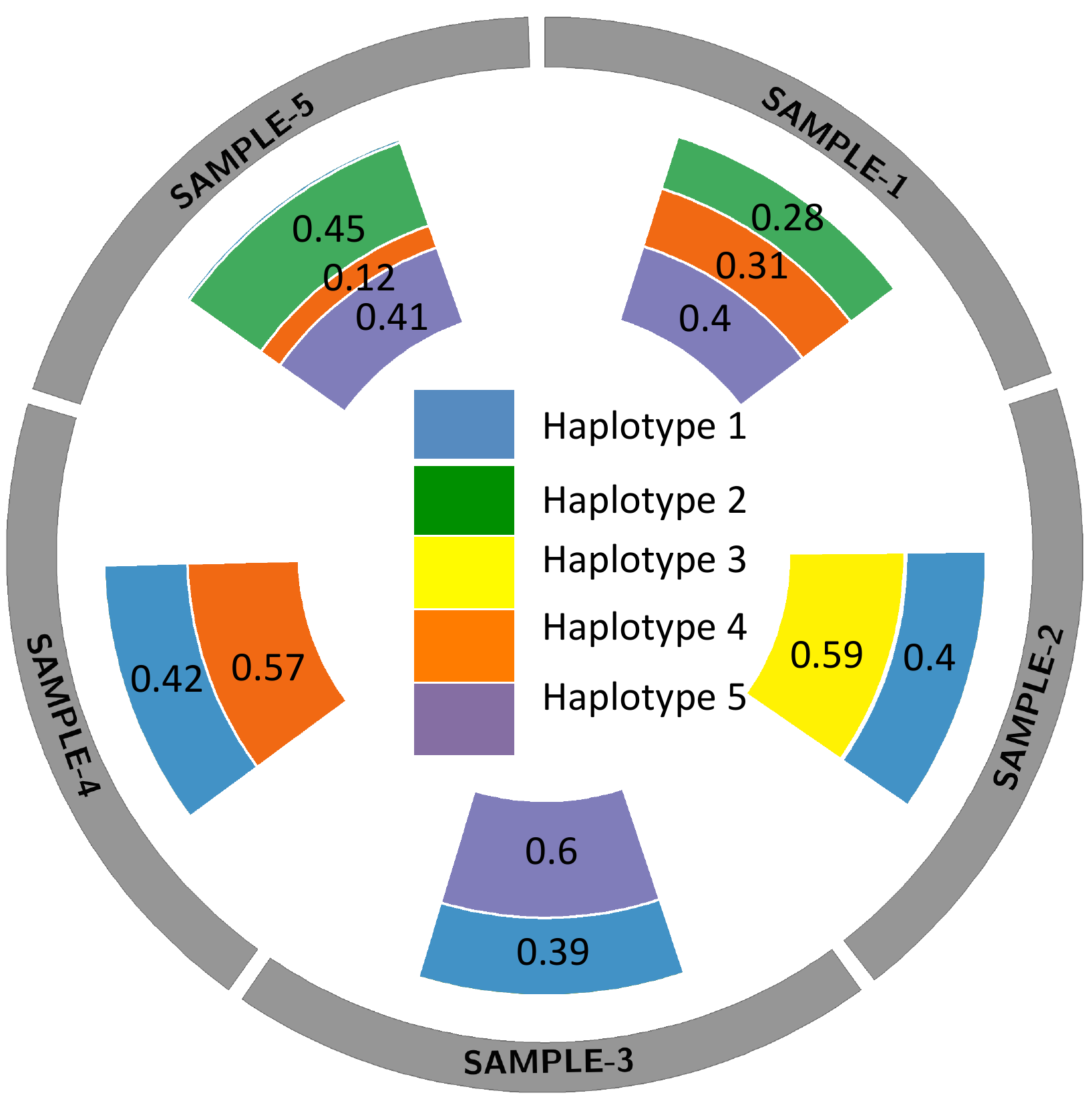} 

\\
 (a) & (b) \\
\end{tabular}
\end{center}
\caption{ Summary of TH analysis using the PDAC data with 118 SNVs. (a):
  Posterior estimated haplotype genotypes (columns) of the selected
  SNVs. Five haplotypes are identified. A green/red block indicates a
  mutant/wildtype allele at the SNV in the haplotype. (b): A circos
  plot showing the estimated
  proportion $\hat{w}_{tc}$ with $\Chat=5$ haplotypes. 
   }

\label{fig:data2}
\end{figure}

Figure \ref{fig:data2} summarizes the haplotype analysis results. In (a) we
present an estimate of $\bZ$, as the point estimate minimizing the objective function $Q$
in our runs. Each of the five columns represents a haplotype and
each row represent an SNV, with green and red blocks indicating
mutant and wildtype  genotypes. In (b) we use a circos plot
\citep{Krzywinski:2009} to present
the proportions of the five estimated haplotypes for each
sample. Samples 1 and 5 possess the same three haplotypes (2, 4, and 5)
while samples 2-4 all possess haplotype 1 and another distinct
haplotype. The results show that most tumor samples consist of only two
haplotypes, except samples 1 \& 5, which have three haplotypes.
No two samples share a complete set of haplotypes, reflecting the
polymorphism between individuals. 
Lastly, haplotypes 1 , 4, and 5  are more prevalent than the other haplotypes, both
appearing in three out of five samples. Haplotype 3 is the least
prevalent appearing only in one sample. 
The corresponding uncertainties, as summarized by $\pbar_{sc}$,
are shown in Figure S6 (a) (see Supplementary Material G). 



For comparisons, we also applied PyClone to infer TH for the same pancreatic cancer data. PyClone identified 27 SNV clusters out of 118 samples, practically rendering the results hard to interpret and biologically less meaningful.

To examine the computational limits of the proposed approach 
we re-analyzed the PDAC data, but now keeping all SNVs that exhibit
significant coverage in all samples and occur in at least two samples
-- not limited to those in KEGG pathways. This filtering left us with
$S=6,599$ SNVs. 

\begin{figure}
\begin{center}
\begin{tabular}{cc}
\includegraphics[width=.35\textwidth]{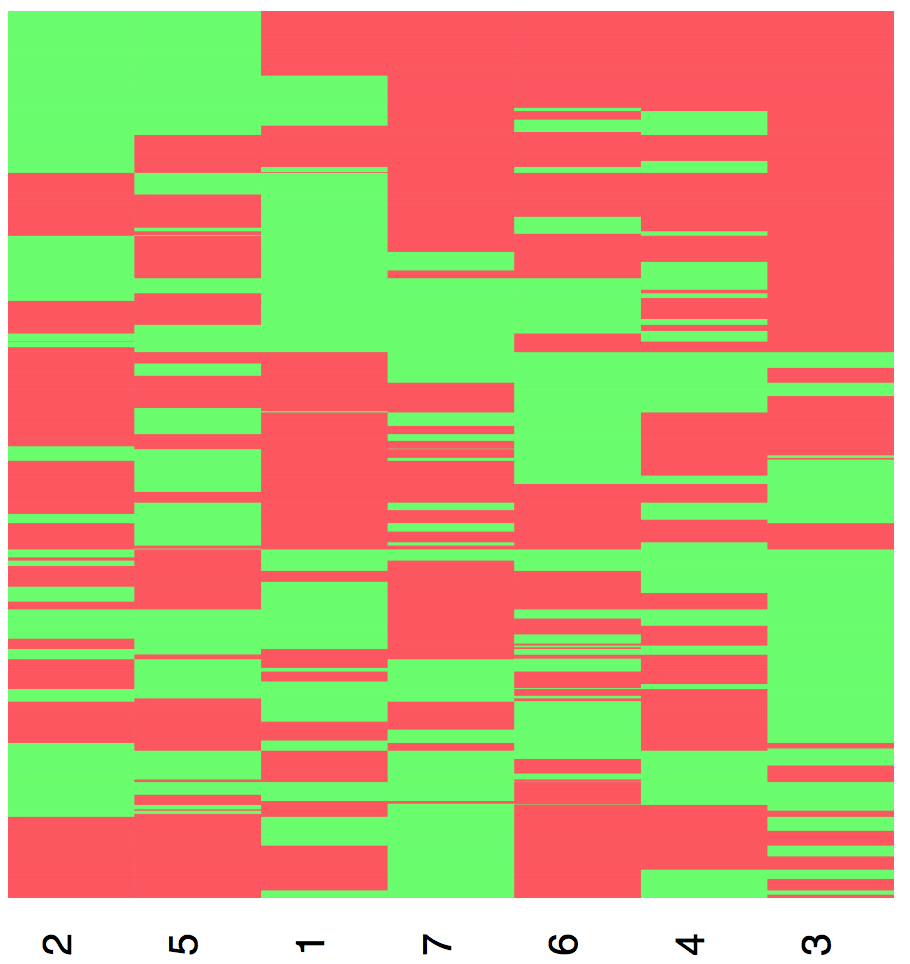}
&
\includegraphics[width=.38\textwidth]{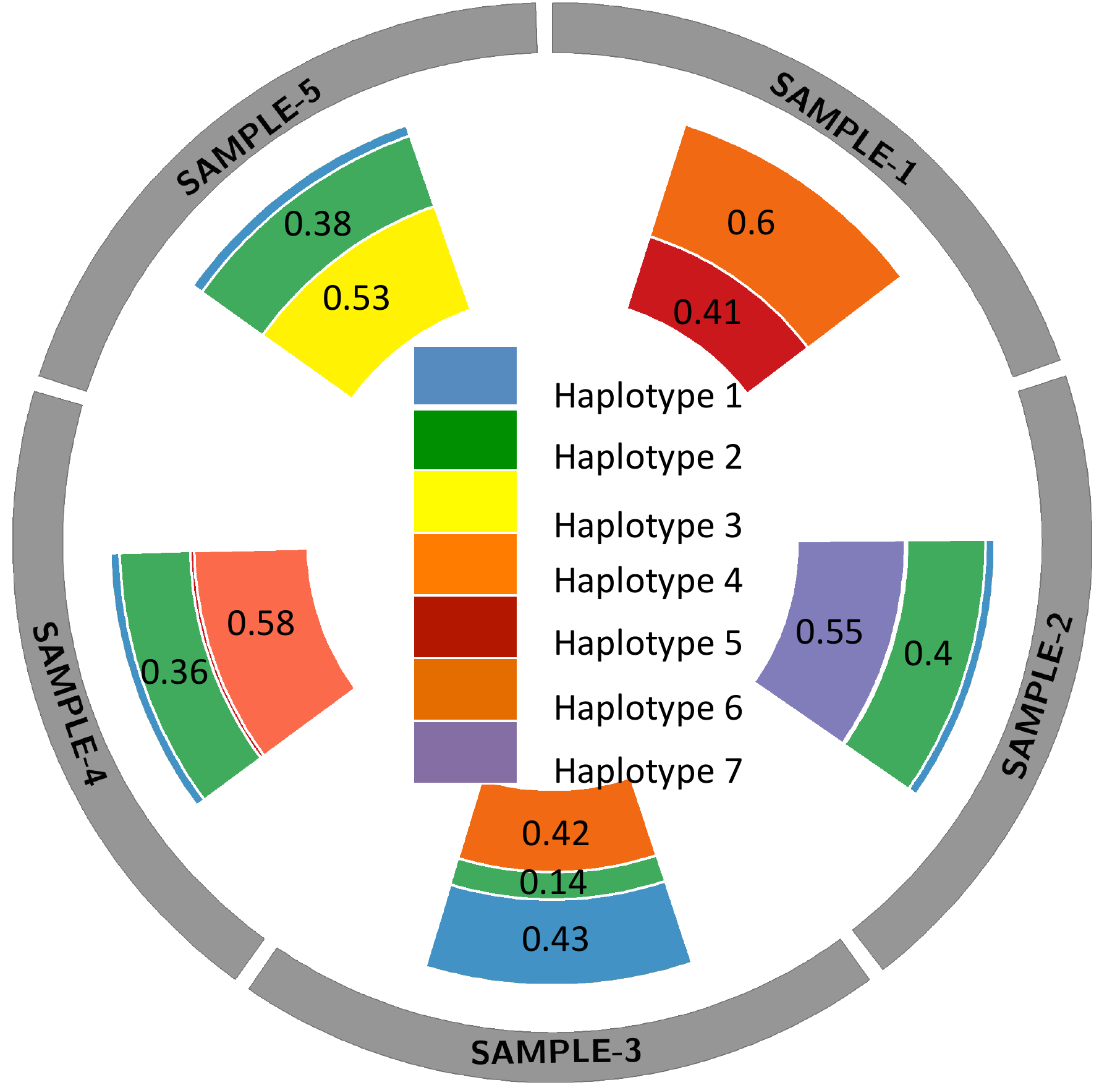} 

\\
 (a) & (b) \\
\end{tabular}
\end{center}
\caption{ Summary of TH analysis using the PDAC data with 6,599 SNVs. (a):
  Posterior estimated haplotype genotypes (columns) of the selected
  SNVs. Seven haplotypes are identified. A green/red block indicates a
  mutant/wildtype allele at the SNV in the haplotype. (b): A circos
  plot showing the estimated
  proportion $\hat{w}_{tc}$ with $\Chat=7$ haplotypes. 
  }

\label{fig:data3}
\end{figure}

We applied the proposed FL-means algorithm
with 1,000 random initializations. Each run of the FL-means algorithm took less than 2 minutes.  After searching for $\lambda^2$
using the suggested heuristic, we estimated $\Chat=7$. The
full inference 
results are summarized in Figure \ref{fig:data3}. Interestingly,
we found patterns that were similar to the previous analysis, such as that each
sample possesses mostly two haplotypes. 
However, with more
SNVs, there are now three (as opposed to one) distinct haplotypes, each only
appearing in one out of five samples. This is not surprising since
with more somatic SNVs, by definition the chance of having common haplotypes between
different tumor samples will reduce. The heatmap in Figure S6 (b) (see Supplementary Material G)
shows the estimated uncertainties of the estimated seven haplotypes. 

Lastly, we applied model \eqref{eq:model2} for subclonal inference
across samples. Examining Figures \ref{fig:data2}(b) and
\ref{fig:data3}(b), we see that most samples (except sample 1 in
Figure \ref{fig:data2}(b)) possess two major haplotypes, which can be
explained by having a single clone, i.e., no subclones. We observed the
same results in subclonal inference: the five samples are clonal, each
possessing a different local genome on the selected SNVs. While the
results are less interesting, they suggest that the proposed 
models do not falsely infer subclonal structure where there is
none.

\section{Conclusion}
\label{sec:con}
We introduce small-variance asymptotics for
the MAP in a feature allocation model  under a binomial likelihood 
as it arises in inference for tumor heterogeneity. 
The proposed FL-algorithm uses a scaled version of the
binomial likelihood that can be introduced as a special case of
scaled exponential family models that are based on writing the
sampling models in terms of Bregman divergences. The algorithm
provides simple and scalable inference to feature allocation problems.


 Inference in the three datasets shows that the proposed 
inference approaches
correctly report subclonal or clonal inference for intra-TH and
inter-patient TH. More importantly, haplotype inference reveals 
shared genotypes on selected SNVs recurring across samples.
Such sharing would provide valuable information for future disease
prognosis, taking advantage of the innovation proposed in this
paper. 
For example, \citet{ding2012clonal} demonstrated that the relapse
of acute myeloid leukemia was associated with new mutations acquired by
a subpopulation of cancer cells derived from the original population.

In this paper we assumed diploidy or two copies of DNA for all
genes. However, the copy number may vary in cancer cells, known as
copy number variants (CNVs). SNVs and CNVs coexist throughout the
genome. CNVs can alter the total read count mapped to a  locus and
eventually the observed VAFs. 
Integrating inference on both, CNVs and SNVs, to infer tumor
heterogeneity could be a useful extension of the proposed model.
Finally, in the proposed approach we implicitly treat normal tissue
like any another subclone. If matching normal samples were available
the model could be extended to borrow strength across tumor and normal
samples by identifying a subpopulation of normal cells as part of the
deconvolution.

\section*{Supplementary Material}
Supplementary material is available under the Paper Information link at the JASA website. 

\section*{Acknowledgment}
Yanxun Xu, Peter M\"{u}ller and Yuan Ji's  research is partially supported by NIH
R01 CA132897.

\bibliographystyle{apalike}
\bibliography{MAD}

\begin{thebibliography}{}

\bibitem[Andor et~al., 2014]{andor2014expands}
Andor, N., Harness, J.~V., M{\"u}ller, S., Mewes, H.~W., and Petritsch, C.
  (2014).
\newblock Expands: expanding ploidy and allele frequency on nested
  subpopulations.
\newblock {\em Bioinformatics}, 30(1):50--60.

\bibitem[Banerjee et~al., 2005]{banerjee2005clustering}
Banerjee, A., Merugu, S., Dhillon, I.~S., and Ghosh, J. (2005).
\newblock Clustering with bregman divergences.
\newblock {\em The Journal of Machine Learning Research}, 6:1705--1749.

\bibitem[Bregman, 1967]{bregman1967relaxation}
Bregman, L.~M. (1967).
\newblock The relaxation method of finding the common point of convex sets and
  its application to the solution of problems in convex programming.
\newblock {\em USSR computational mathematics and mathematical physics},
  7(3):200--217.

\bibitem[Broderick et~al., 2012a]{broderick2012beta}
Broderick, T., Jordan, M.~I., and Pitman, J. (2012a).
\newblock Beta processes, stick-breaking and power laws.
\newblock {\em Bayesian analysis}, 7(2):439--476.

\bibitem[Broderick et~al., 2012b]{broderick2012mad}
Broderick, T., Kulis, B., and Jordan, M.~I. (2012b).
\newblock Mad-bayes: Map-based asymptotic derivations from bayes.
\newblock {\em arXiv preprint arXiv:1212.2126}.

\bibitem[Dempster et~al., 1977]{dempster1977maximum}
Dempster, A.~P., Laird, N.~M., and Rubin, D.~B. (1977).
\newblock Maximum likelihood from incomplete data via the em algorithm.
\newblock {\em Journal of the Royal Statistical Society. Series B
  (Methodological)}, pages 1--38.

\bibitem[Ding et~al., 2012]{ding2012clonal}
Ding, L., Ley, T.~J., Larson, D.~E., Miller, C.~A., Koboldt, D.~C., Welch,
  J.~S., Ritchey, J.~K., Young, M.~A., Lamprecht, T., McLellan, M.~D., et~al.
  (2012).
\newblock Clonal evolution in relapsed acute myeloid leukaemia revealed by
  whole-genome sequencing.
\newblock {\em Nature}, 481(7382):506--510.

\bibitem[Gerlinger et~al., 2012]{gerlinger2012intratumor}
Gerlinger, M., Rowan, A.~J., Horswell, S., Larkin, J., Endesfelder, D.,
  Gronroos, E., Martinez, P., Matthews, N., Stewart, A., Tarpey, P., et~al.
  (2012).
\newblock Intratumor heterogeneity and branched evolution revealed by
  multiregion sequencing.
\newblock {\em New England Journal of Medicine}, 366(10):883--892.

\bibitem[Ghoshal, 2010]{Ghoshal:10}
Ghoshal, S. (2010).
\newblock The {D}irichlet process, related priors and posterior asymptotics.
\newblock In Nils Lid~Hjort, Chris~Holmes, P.~M. and Walker, S.~G., editors,
  {\em Bayesian Nonparametrics}, pages 22--34. Cambridge University Press.

\bibitem[Green, 1995]{green1995reversible}
Green, P.~J. (1995).
\newblock Reversible jump markov chain monte carlo computation and bayesian
  model determination.
\newblock {\em Biometrika}, 82(4):711--732.

\bibitem[Griffiths and Ghahramani, 2006]{griffiths2005infinite}
Griffiths, T.~L. and Ghahramani, Z. (2006).
\newblock {Infinite latent feature models and the Indian buffet process}.
\newblock In Weiss, Y., Sch\"olkopf, B., and Platt, J., editors, {\em Advances
  in Neural Information Processing Systems 18}, pages 475--482. MIT Press,
  Cambridge, MA.

\bibitem[Hartigan and Wong, 1979]{hartigan1979algorithm}
Hartigan, J.~A. and Wong, M.~A. (1979).
\newblock Algorithm as 136: A k-means clustering algorithm.
\newblock {\em Journal of the Royal Statistical Society. Series C (Applied
  Statistics)}, 28(1):100--108.

\bibitem[Hastie et~al., 2001]{trevor2001elements}
Hastie, T., Tibshirani, R., and Friedman, J. J.~H. (2001).
\newblock {\em The elements of statistical learning}, volume~1.
\newblock Springer New York.

\bibitem[Kanehisa et~al., 2010]{kanehisa2010kegg}
Kanehisa, M., Goto, S., Furumichi, M., Tanabe, M., and Hirakawa, M. (2010).
\newblock Kegg for representation and analysis of molecular networks involving
  diseases and drugs.
\newblock {\em Nucleic acids research}, 38(suppl 1):D355--D360.

\bibitem[Karlof, 2005]{karlof2005integer}
Karlof, J.~K. (2005).
\newblock {\em Integer programming: theory and practice}.
\newblock CRC Press.

\bibitem[Krzywinski et~al., 2009]{Krzywinski:2009}
Krzywinski, M., Schein, J., Birol, I., Connors, J., Gascoyne, R., Horsman, D.,
  Jones, S., and Marra, M. (2009).
\newblock Circos: an information aesthetic for comparative genomics.
\newblock {\em Genome Res.}, 19(9):1639--45.

\bibitem[Kulis and Jordan, 2011]{kulis2011revisiting}
Kulis, B. and Jordan, M.~I. (2011).
\newblock Revisiting k-means: New algorithms via bayesian nonparametrics.
\newblock {\em arXiv preprint arXiv:1111.0352}.

\bibitem[Landau et~al., 2013]{landau2013evolution}
Landau, D.~A., Carter, S.~L., Stojanov, P., McKenna, A., Stevenson, K.,
  Lawrence, M.~S., Sougnez, C., Stewart, C., Sivachenko, A., Wang, L., et~al.
  (2013).
\newblock Evolution and impact of subclonal mutations in chronic lymphocytic
  leukemia.
\newblock {\em Cell}, 152(4):714--726.

\bibitem[Larson and Fridley, 2013]{larson2013purbayes}
Larson, N.~B. and Fridley, B.~L. (2013).
\newblock Purbayes: estimating tumor cellularity and subclonality in
  next-generation sequencing data.
\newblock {\em Bioinformatics}, 29(15):1888--1889.

\bibitem[Lee et~al., 2013]{juhee2013feature}
Lee, J., M\"uller, P., Ji, Y., and Gulukota, K. (2013).
\newblock A feature allocation model for tumor heterogeneity.
\newblock {\em Technical Report}.

\bibitem[Liu, 2008]{liu2008monte}
Liu, J.~S. (2008).
\newblock {\em Monte Carlo strategies in scientific computing}.
\newblock springer.

\bibitem[{McKenna} et~al., 2010]{mckennathe2010}
{McKenna}, A., Hanna, M., Banks, E., Sivachenko, A., Cibulskis, K., Kernytsky,
  A., Garimella, K., Altshuler, D., Gabriel, S., Daly, M., and Mark, D. (2010).
\newblock The genome analysis toolkit: a {MapReduce} framework for analyzing
  next-generation {DNA} sequencing data.
\newblock {\em Genome research}, 20(9):1297--1303.

\bibitem[Roth et~al., 2014]{roth2014pyclone}
Roth, A., Khattra, J., Yap, D., Wan, A., Laks, E., Biele, J., Ha, G., Aparicio,
  S., Bouchard-C{\^o}t{\'e}, A., and Shah, S.~P. (2014).
\newblock Pyclone: statistical inference of clonal population structure in
  cancer.
\newblock {\em Nature methods}.

\bibitem[Teh et~al., 2007]{teh2007stick}
Teh, Y.~W., G{\"o}r{\"u}r, D., and Ghahramani, Z. (2007).
\newblock Stick-breaking construction for the indian buffet process.
\newblock In {\em International Conference on Artificial Intelligence and
  Statistics}, pages 556--563.

\bibitem[Thibaux and Jordan, 2007]{thibeauxjordan:07}
Thibaux, R. and Jordan, M. (2007).
\newblock Hierarchical beta processes and the indian buffet process.
\newblock In {\em Proceedings of the 11th Conference on Artificial Intelligence
  and Statistics (AISTAT)}, Puerto Rico.

\end{thebibliography}
\clearpage


\end{document}